\documentclass[letterpaper,10pt,twocolumn]{IEEEtran}
\usepackage{graphics} 
\usepackage{graphicx}
\usepackage{epsfig}
\usepackage{subfigure}
\usepackage{amssymb}
\usepackage{amsbsy}
\usepackage{amsmath}
\usepackage{cite}
\usepackage{url}
\usepackage{mathptmx} 

\usepackage{color}

\usepackage{placeins}
\usepackage{tabularx,colortbl}
\usepackage{times,amsmath,epsfig}
\usepackage{xspace,latexsym,syntonly}
\usepackage{amssymb}
\usepackage{amsfonts}
\usepackage{textcomp}
\usepackage{subfigure}
\usepackage{amsbsy}
\usepackage{cite}
\usepackage{verbatim}

\usepackage{framed}

\usepackage{epstopdf}
\usepackage{bbm,bm}
\usepackage{multirow}
\usepackage{mathtools}
\usepackage{hyperref} \hypersetup{pdfborder={0 0 0}, colorlinks=false,linkcolor=black}

\usepackage{algorithm} 
\usepackage{algorithmicx} 
\usepackage{algpseudocode}

\newcommand{\cD}{{\cal D}}

\newtheorem{theorem}{Theorem}
\newtheorem{lemma}{Lemma}

\newtheorem{definition}{Definition}

\newcommand{\beq}{\begin{equation}}
\newcommand{\eeq}{\end{equation}}
\newcommand{\bea}{\begin{array}}
\newcommand{\ena}{\end{array}}
\newcommand{\bds}{\begin {itemize}}
\newcommand{\eds}{\end {itemize}}
\newcommand{\bdf}{\begin{definition}}
\newcommand{\blm}{\begin{lemma}}
\newcommand{\edf}{\end{definition}}
\newcommand{\elm}{\end{lemma}}
\newcommand{\bthm}{\begin{theorem}}
\newcommand{\ethm}{\end{theorem}}
\newcommand{\bprp}{\begin{prop}}
\newcommand{\eprp}{\end{prop}}
\newcommand{\bcl}{\begin{claim}}
\newcommand{\ecl}{\end{claim}}
\newcommand{\bcr}{\begin{coro}}
\newcommand{\ecr}{\end{coro}}
\newcommand{\bquest}{\begin{question}}
\newcommand{\equest}{\end{question}}

\newcommand{\larrow}{{\larrow}}

\def\urltilda{\kern -.15em\lower .7ex\hbox{\~{}}\kern .04em}

\begin{document}\title{Optimal Nested Test Plan for \\Combinatorial Quantitative Group Testing}
\author{Chao Wang, Qing Zhao, Chen-Nee Chuah
\thanks{Chao Wang and Qing Zhao are with the School of Electrical and Computer Engineering, Cornell University. Emails: \{cw733, qz16\}@cornell.edu.}
\thanks{Chen-Nee Chuah is with the Department of Electrical and Computer Engineering, University of California, Davis. Email: chuah@ucdavis.edu.}
\thanks{This work was supported by the National Science Foundation under Grants CNS-1321115 and CCF-1320065 and by the U.S. Army Research Office under Grant W911NF-17-1-0464. This paper was presented in part at the 2015 Information Theory and Applications Workshop.}
}
\date{}
\maketitle

%
\begin{abstract}
\label{sec:abstract}
We consider the quantitative group testing problem where the objective is to identify defective items in a given population based on results of tests performed on subsets of the population. Under the quantitative group testing model, the result of each test reveals the number of defective items in the tested group. The minimum number of tests achievable by nested test plans was established by Aigner and Schughart in 1985 within a minimax framework. The optimal nested test plan offering this performance, however, was not obtained. In this work, we establish the optimal nested test plan in closed form. This optimal nested test plan is also order optimal among all test plans as the population size approaches infinity. Using heavy-hitter detection as a case study, we show via simulation examples orders of magnitude improvement of the group testing approach over two prevailing sampling-based approaches in detection accuracy and counter consumption. Other applications include anomaly detection and wideband spectrum sensing in cognitive radio systems. 
\end{abstract}
%
\def\keywords{\vspace{.5em}
{\bfseries\textit{Index Terms}---\,\relax%
}}
\def\endkeywords{\par}
\keywords
Group testing, adaptive test plan, heavy hitter detection, anomaly detection, traffic measurements, spectrum sensing.
\section{Introduction}
\label{sec:intro}

\subsection{Classic Group Testing}
\label{ssec:GT}
The group testing problem is concerned with identifying defective items in a given population by performing tests over subsets of the population.  The objective is a test plan with a minimal number of tests identifies all defective items.\par
Under the classic model, each test gives a binary result, 
indicating whether the tested group contains any defective items.
The problem was first motivated by the practice of screening draftees with syphilis during World War II, and the idea of testing pooled blood samples from a group of people (rather than testing each person one by one) was initiated by Robert Dorfman~\cite{Dorfman}.

There are two formulations of the group testing problem, known as \emph{probabilistic group testing} (PGT) and \emph{combinatorial group testing} (CGT). The former is a Bayesian formulation that assumes a probabilistic model on the defective items and aims to minimize the expected number of tests for identifying all defective items~\cite{sobel1959}. The latter is a minimax formulation that assumes a deterministic value $d$ for the total number of defective items and aims to minimize the number of tests in the worst case (among all compositions of the defective set of size $d$)~\cite{Li,Katona,Du_Hwang_book}. 

Under both formulations, the test plans can be adaptive or non-adaptive. Adaptive test plans are sequential in nature: which group to test next depends on the outcome of the previous tests. The studies in \cite{sobel1959, Li, Katona} mentioned above all focus on adaptive test plans. Non-adaptive group testing is a one-stage problem in which all actions are determined before any test is performed. Non-adaptive test plans are often represented by matrices~\cite{Ngo, Du_Hwang_book_nonadp}.

The classic group testing problem has seen a wide range of applications, including mutiaccess communications~\cite{JKWolf, wolf1985principles, Berger}, idle channel detection in the radio spectrum~\cite{Murthy}, compressed sensing~\cite{Cheraghchi_comp}, network tomography~\cite{Cheraghchi_graph}, and anomaly detection~\cite{Thai, Khattab}. In particular, non-adaptive group testing has been widely applied to DNA sequencing and DNA library screening ~\cite{Ngo, Balding}.

\subsection{Quantitative Group Testing}
\label{ssec:quantitative}
In a quantitative group testing problem, a test reveals the number of defective items in the tested group, a finer observation model than the binary model assumed in the classic group testing~\cite{Du_Hwang_book}. It is also known as the coin weighing problem with a spring scale first introduced by Shapiro in 1960~\cite{Problem1399}. The problem is to identify $d$ counterfeit coins in a collection of $n$ coins. The weights of the authentic and counterfeit coins are known. Thus each weighing gives the number of counterfeit coins in the tested group. \par 

Most studies on quantitative group testing focus on non-adaptive test plans, see, for example,~\cite{F60,L75,ER63} on the case of unknown $d$ and~\cite{D75,CK08} on the case of known $d$. Adaptive test plans have been studied mostly for the special case of $d=2$ (see~\cite{A86,Hao90,GMSV92}). Only a couple of results are available on adaptive test plans for the general case of $0<d<n$. In particular, Aigner and Schughart considered a class of adaptive test plans with a \emph{nested} structure~\cite{AS85}. Specifically, in a nested test plan, once a test reveals a group containing defective items, the next test must be a proper subset of this group. They established the performance (i.e., the number of required tests) of the optimal nested test plan. The optimal nested test plan itself, however, was not obtained. In~\cite{B09}, Bshouty developed a semi-adaptive test plan that integrates a bisecting search with a non-adaptive test plan. It was shown that this semi-adaptive test plan can be constructed in polynomial time and has a performance no worse than twice of the information-theoretic lower bound. However, the algorithm may fail to construct a valid test plan in certain cases\footnote{One such example is when $n=200$ and $d=52$, the algorithm fails to construct the corresponding $(3,3,4,5,6,7,8,8,9,9)$-\emph{Detection Matrix}.}.

The applications of quantitative group testing include the uniquely decodable codes for the noiseless $n$-user adder channel problem\cite{CW79}, and the construction of unknown graphs from additive queries\cite{A86,CK08}. Several variations of the problem can be found in~\cite{BM10,GK97,Han&Frazier&Jedynak:14}.

\subsection{Main Results and Applications}
\label{ssec:related}

In this paper, we consider the quantitative group testing problem under the CGT formulation with adaptive test plan for both known and unknown $d$. As mentioned above, this problem with known $d$ was first studied by Aigner and Schughart in \cite{AS85} in which they established the number of tests required by the optimal nested test plan for identifying $d$ defective items in a population of size $n$. To our best knowledge, the optimal nested test plan remains open. In this paper, we obtain the optimal nested test plan in closed form. \par

The optimal number of tests $N(n,d)$ was given in \cite{AS85} in the form of inequalities. From these inequalities, we obtain a closed-form expression of $N(n,d)$. We also show that the sequence of $N(n,d)$ in $n$ for fixed $d$ has a clean pattern which can be illustrated in a \emph{frame-segment} structure. However, since $N(n,d)$ is a nonlinear integer-valued function involving multiple layered ceiling functions, directly obtaining the optimal test plan from $N(n,d)$ by solving an integer optimization problem is intractable. Our approach is to first establish three key properties of $N(n,d)$ and of the optimal test plan. Based on these properties, we obtain the optimal test plan in closed form using induction, which also has a clean frame-segment structure corresponding to the pattern of $N(n,d)$. We point out that establishing these properties of $N(n,d)$ itself is nontrivial due to the complex nonlinearity of $N(n,d)$ in both $n$ and $d$.  \par

We then focus on the application of heavy hitter detection for traffic monitoring and anomaly detection in the Internet and other communication networks. For Internet traffic, it is a common observation that a small percentage of high-volume flows (referred to as heavy hitters) account for most of the total traffic\cite{Thompson_wide}. In particular, it was shown in~\cite{Fang} that the top (in terms of volume) $9\%$ of flows make up $90.7\%$ of the total traffic over the Internet. Quickly identifying the heavy hitters is thus crucial to network stability and security. However, the large number of Internet flows makes individual monitoring extremely inefficient if not impossible.
A quantitative group testing approach to heavy hitter detection offers an efficient solution under which the number of required measurements for reliable detection grows logarithmically rather than linearly with the number of flows. Indeed, recent advances in software defined networking
(SDN) allow programmable routers to count aggregated flows that match a given IP prefix\cite{Yu2013SDN}.\par

The quantitative group testing model stems from the fact that the difference between the average traffic rates of heavy hitters and normal flows is large, which allows for accurate estimation of the number of heavy hitters from random measurements of the aggregated traffic load. Through simulation examples, we examine the performance of the group testing approach in terms of detection delay, detection accuracy, and counter consumption. Significant improvement over two prevailing sampling-based approach is observed.\par

Other potential applications include detecting idle channels in the radio spectrum when the signal strength is relatively even across busy channels and much higher than the noise level in idle channels (the high SNR regime).

\subsection{Discussion of Related Work}
\label{ssec:add_related}

Much of the related work has been discussed in the preceding subsections. Here we provide additional related work, focusing on the comparison between adaptive and non-adaptive group testing approaches and the connection between quantitative group testing and compressed sensing.

\subsubsection{Adaptive vs. Non-Adaptive Group Testing}
Most work on group testing focuses on non-adaptive test plans. A non-adaptive test plan can be represented by a binary measurement matrix with columns corresponding to items, rows corresponding to tests, and the $(i,j)$th element indicating whether item $j$ is included in the $i$th group test. Constructing the measurement matrix can be cast as a source coding problem, and the superimposed code and the uniquely decipherable code have been used in developing non-adaptive test plans (see, for example, \cite{Kautz1964nonrandom, Berger2002asymptotic, Gilbert2008group, Emad2014semiquantitative, Erdos1985families}). It is this connection to source coding that brings mathematical tractability to non-adaptive group testing, a treat seldom enjoyed by adaptive group testing. Allowing parallel implementation with all tests run simultaneously also makes non-adaptive test plans attractive in applications that involve a lengthy delay in obtaining test results. The disadvantages of non-adpative test plans lie in the computational complexity of the coding/decoding processes, high storage requirement, and difficulty to adjust to cases with unknown or time-varying population compositions~$\{n, d\}$. \par

Adaptive test plans, in contrast, are more suitable for online applications where the values of $n$ and/or $d$ are not prefixed. Furthermore, the optimal nested test plan developed in this work is given in closed form and has a clean frame-segment structure; little offline or online computation is needed. The inherent tree structure of the nested test plan also leads to low memory requirement. It is thus particularly attractive for online applications such as real-time heavy hitter detection where $n$ and $d$ are not prefixed and computational, memory, and counter resources are stringent.

\subsubsection{Connection with Compressed Sensing}

The quantitative group testing problem shares similarity with the compressed sensing problem. In compressed sensing, the objective is to recover a sparse signal from linear measurements. Specifically, given an $n$-dimensional sparse signal with a support size $d$, the goal is to identify the support set (non-zero elements of the signal) with a minimum number of projections. The differences between compressed sensing and quantitative group testing are in the signal model and constraints on the measurement/projection matrix. Most work on compressed sensing assumes real-valued signals and allow real-valued measurement matrices. Quantitative group testing, when viewed as compressed sensing, deals with binary signals that are not necessarily sparse and require the measurement matrix to be binary valued. There are a number of non-adaptive compressed sensing algorithms in the literature that result in binary-valued measurement matrices (see, for example,~\cite{Xu2007_efficient, Xu2007_further, Indyk2008_near_L1, Jafarpour2009_efficient, Berinde2008_combining, cheraghchi2010derandomization}). However, in addition to the sparsity requirement, these non-adaptive strategies suffer the same difficulties as non-adaptive group testing algorithms in online applications as discussed above. \par

Several adaptive compressed sensing algorithms exist in the literature~\cite{Malloy2014_near_opt_adp, Iwen2012adaptive,Haupt2012sequentially,Ji2008bayesian}. 
They were shown to outperform non-adaptive algorithms in sample complexity and detection performance. Most of the adaptive compressed sensing algorithms, however, are not directly applicable to quantitative group testing due to the real-valued measurement matrices.
The only exceptions are~\cite{Malloy2014_near_opt_adp, Iwen2012adaptive}, in which two similar bisecting search approaches were introduced. While the problems formulated in~\cite{Malloy2014_near_opt_adp, Iwen2012adaptive} were to estimate a real-valued sparse signal under certain constraints, the bisecting search approach proposed there can be applied to the quantitative group testing problem and constitutes a suboptimal nested test plan. In this work, we develop the optimal nested test plan for combinatorial quantitative group testing.

\section{Problem Formulation}
\label{sec:formulation}

Consider a population of $n$ items. It is known that among these $n$ items, $d$ are defective (the issue of unknown $d$ is addressed in Section~\ref{ssec:unknown}). Let $(n,d)$ denote the corresponding quantitative CGT problem. We assume that $1\le d\le n-1$ to avoid the trivial scenarios of $d=0$ and $d=n$. \par

For a given $(n,d)$, an adaptive test plan $\pi$ is a sequence of decision rules $\{\pi_1,\pi_2,\ldots\}$ where $\pi_t$ maps from the outcomes of the previous $t-1$ tests to the subset of items to be tested in the $t$th test. With a slight abuse of notation, $\pi_t$~is also used to denote the subset of items tested in the $t$th test under test plan $\pi$. Let $N_{\pi}(n,d; \cD)$ denote the number of tests required for identifying all $d$ defective items under $\pi$ when the $d$ defective items are specified by the set $\cD$. Note that $n$ and $d$ are known while $\cD$ is unknown and is what the test plan needs to identify. Under the combinatorial formulation, the performance of a test plan is determined by the worst instance of $\cD$ among all subsets with size $d$. The performance of $\pi$, denoted by $N_{\pi}(n,d)$, is thus given by

\setcounter{equation}{0}
\begin{equation}\label{CGT_d}
N_{\pi}(n,d)=\max_{\cD\subset [n], |\cD|=d} N_{\pi}(n,d; \cD),
\end{equation}
where $[n]$ denotes the set of all $n$ items.

In this work, we focus on a family of test plans that exhibit a tree structure. This family is referred to as the nested test plan as defined below. 

\begin{definition}
An adaptive test plan $\pi=\{\pi_1,\pi_2,\ldots\}$ is a nested test plan if for all $t\ge 1$ and $k\ge 1$, the tested groups $\pi_t$ and $\pi_{t+k}$ at the $t$th and the $(t+k)$th tests satisfy either $\pi_t \cap \pi_{t+k} =\pi_{t+k}$ or $\pi_t \cap \pi_{t+k} =\emptyset$.
\end{definition}

Based on the above definition, it is not difficult to show that for every instance of $(n,d; \mathcal{D})$, the $N_\pi(n,d;\mathcal{D})$ tested groups $\{\pi_1,\pi_2,\ldots, \pi_{N_\pi(n,d;\mathcal{D})}\}$ form a tree. Specifically, consider a graph with $N_\pi(n,d;\mathcal{D})$ nodes representing each of the tested groups and a root node representing the set $[n]$ of the entire population. An edge exists between two nodes if and only if one of them is the smallest superset of the other. It can be shown such a graph resulting from a nested test plan is acyclic.

Our objective is an optimal nested test plan $\pi^*$ given by
\begin{equation}\label{CGT_opt}
\pi^*=\arg\min_{\pi\in \Pi} N_{\pi}(n,d),
\end{equation}
where $\Pi$ denotes the family of all nested test plans. To simplify the notation, the performance of the optimal nested test plan $\pi^*$ is denoted by $N(n,d)$ (rather than $N_{\pi^*}(n,d)$).\par

For a given CGT $(n,d)$, due to the symmetry among items, the worst-case performance $N_\pi(n,d)$ of any test plan $\pi$ depends on the first test only through the size of the tested group but not the specific composition of the group. Suppose that the first test consists of $m$ items and the outcome reveals that $d_1$ items among these $m$ are defective. For a nested test plan, this first test decomposes the original CGT problem of $(n,d)$ into two independent CGT problems of $(m,d_1)$ and $(n-m,d-d_1)$. Obviously, $d_1$ cannot exceeds $m$ or $d$. At the same time, $d_1$ cannot be smaller than $0$ or $d-(n-m)$ (the latter is due to the fact that the $n-m$ untested items consist of at most $(n-m)$ defective items). Combined with the minimax nature of the CGT formulation, this leads to the following recursive equation for $N(n,d)$:
\begin{equation}\label{opt}
N(n,d) = 1 + \min_m \max_{d_1} \big\lbrace N(m,d_1) + N(n-m,d-d_1)\big\rbrace,
\end{equation}
where the maximization over $d_1$ is among integers in the range of $\max\lbrace 0, d+m-n \rbrace $ to $\min \lbrace m, d \rbrace$ and the minimization over $m$ can be set to integers $1,2,\ldots, \left\lfloor \frac{n}{2}\right\rfloor$ (since testing a group of size $m$ is equivalent to testing a group of size $n-m$). 

Due to this decomposition of the problem into two independent problems of smaller sizes, a nested test plan is fully specified once the first test is determined for all possible population sizes $n$ and all possible numbers  $d$ of defective items. Furthermore, since the composition of the tested group is inconsequential, specifying the size $m$ of the first group test for all $n$ and $d$ suffices. Let $M(n,d)$ denote the value of $m$ that achieves $N(n,d)$ in~\eqref{opt}, i.e.,
\begin{equation}\label{optM}
M(n,d) =   \min_m \max_{d_1} \big\lbrace N(m,d_1) + N(n-m,d-d_1)\big\rbrace.
\end{equation}
The values of $M(n,d)$ for all $n\ge 1$ and $0\le d\le n$ specify the optimal nested test plan for all CGT problems. We point out that when there are multiple values of the group size $m$ that achieve the minimum in \eqref{optM}, $M(n,d)$ is set to the smallest such value. A smaller group size is often preferred in practical applications. \par

The focus on nested test plan is motivated by its analytical tractability, its simple implementation, and its order optimality. Without imposing any structure, the optimal test plan is analytically intractable in general. Obtaining the optimal test plan numerically through exhaustive search is computationally prohibitive due to the combinatorial nature of the problem. The nested structure, however, leads to the clean recursive formulas in (\ref{opt}, \ref{optM}), offering the possibility of explicit analytical characterizations. Nested test plans also enjoy simpler implementation due to the tree-structured splitting of previously tested groups. This tree structure results in lower memory requirement for storing all past test outcomes. It also allows maintaining a certain contiguous property in each tested group, which is often desirable in practice. For example, for the application of heavy hitter detection, the contiguous property is in terms of all flows in the tested group sharing a common IP prefix, which simplifies the router configuration for packet count of the aggregated flow. For the application of spectrum sensing, the contiguous property is in terms of adjacency in the spectrum, which eases filter implementation. Lastly,  the optimal nested test plan is order optimal among all test plans as shown in Section~\ref{ssec:order_opt}.

\section{The Optimal Nested Test Plan}
\label{sec:optimal_test} 
In this section, we establish the optimal nested test plan in closed-form. This result hinges on a compact closed-form expression of $N(n,d)$ and its geometric block-constant structure as established in Lemma~\ref{lemma:Nnd} below.

\subsection{$N(n,d)$ and Its Geometric Block-Constant Structure}
\label{ssec:performance}

In QGT, a test outcome reveals the number of defective items, thus also the number of non-defective items. This symmetry between defective and non-defective items readily leads to $N(n,d) = N(n,n-d)$. It thus suffices to assume $d\leq \left\lfloor \frac{n}{2}\right\rfloor$ unless otherwise noted.

The performance of the optimal nested tested plan is given in the following lemma. 
\begin{lemma}\label{lemma:Nnd}
For a CGT problem $(n,d)$ with $d\le\frac{n}{2}$, we have
\begin{eqnarray}
&&\label{eqN} N(n,d)=(l+1)d+k-1,
\end{eqnarray}
where
\begin{eqnarray}
&&\label{eql}l=\left\lceil\log_2{(n/d)}\right\rceil-1,\\
&&\label{eqk}k=\lceil n/2^l\rceil-d.
\end{eqnarray}
\end{lemma}
\begin{IEEEproof}
The proof is based on the characterization of $N(n,d)$ given in~\cite{AS85} in the form of the following three inequalities.
\begin{eqnarray}
&&\label{ineq1}N(2d,d)\geq 2d-1,\\
&&\label{ineq2}N((d+i)2^{t-1},d)\leq td+i-1,\\
&&\label{ineq3}N((d+i)2^{t-1}+1,d)\geq td+i,
\end{eqnarray}
where $t\geq 2,\ d\geq 1,\ 0\leq i\leq d-1$. Detailed of the proof are given in Appendix~\ref{app:lemma1}. 
\end{IEEEproof}
\medskip

Lemma~\ref{lemma:Nnd} reveals an interleaved block-constant structure with geometrically growing block length of rate $2$. As illustrated in Table~\ref{patternN}, the sequence of $N(n,d)$ in terms of $n$ for a fixed $d$ consists of \emph{frames}, with each frame containing $d$ \emph{segments}. The two positive integers $l$ and $k$ given in \eqref{eql} and \eqref{eqk} are, respectively, the frame index and the segment index. Specifically, each sequence $N(n,d)$ starts at $n=2d$ with $N(2d,d)=2d-1$~(recall that it is sufficient to consider $d\leq \left\lfloor\frac{n}{2}\right\rfloor$). Following this initial value, the rest of the sequence is partitioned into frames with the frame length doubled from one frame to the next. Each frame consists of $d$ segments of equal length with a segment length of $2^l$ in the $l$th frame ($l=1,2,\ldots$). The value of $N(n,d)$ is the same within a segment and increases by $1$ from one segment to the next. \par

We point out that while $N(n,d)$ was determined in~\cite{AS85}, it was specified through the three inequalities given in~(\ref{ineq1}-\ref{ineq3}). The expression given in Lemma~\ref{lemma:Nnd} is not only more compact but also reveals the frame-segment structure of $N(n,d)$. As shown in Section~\ref{sec:proof_thm}, this frame-segment structure of $N(n,d)$ is the key to establishing the optimal nested test plan.

\begin{table}
\centering
\caption{\label{patternN}The Frame-Segment Structure of $N(n,d)$}
\begin{tabular}{l}
\hline
\hline\\
\includegraphics[scale=0.23]{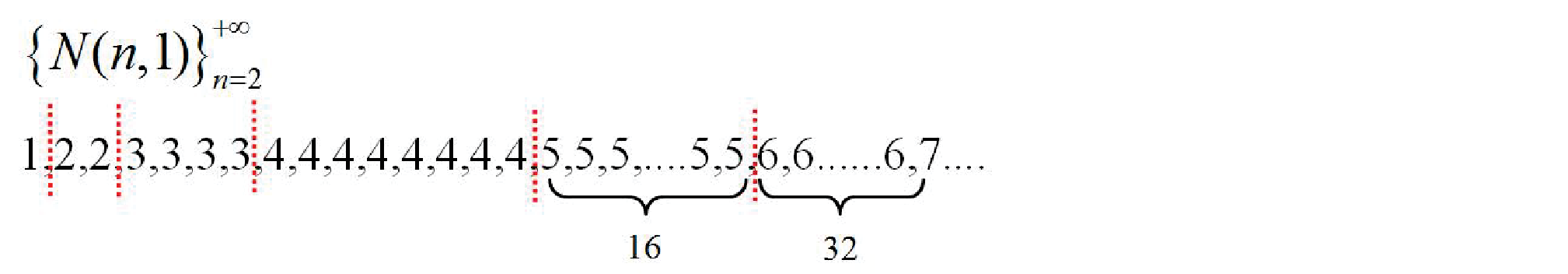}\\
\includegraphics[scale=0.23]{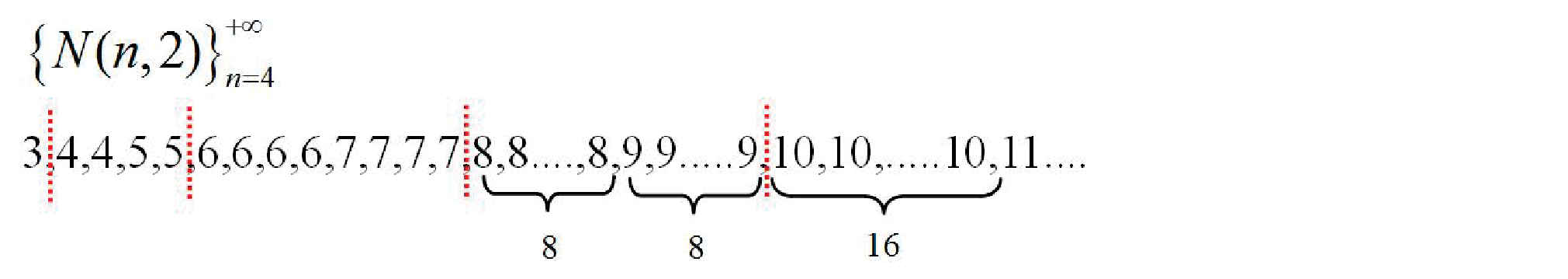}\\
$\quad\quad\vdots$\\
\includegraphics[scale=0.23]{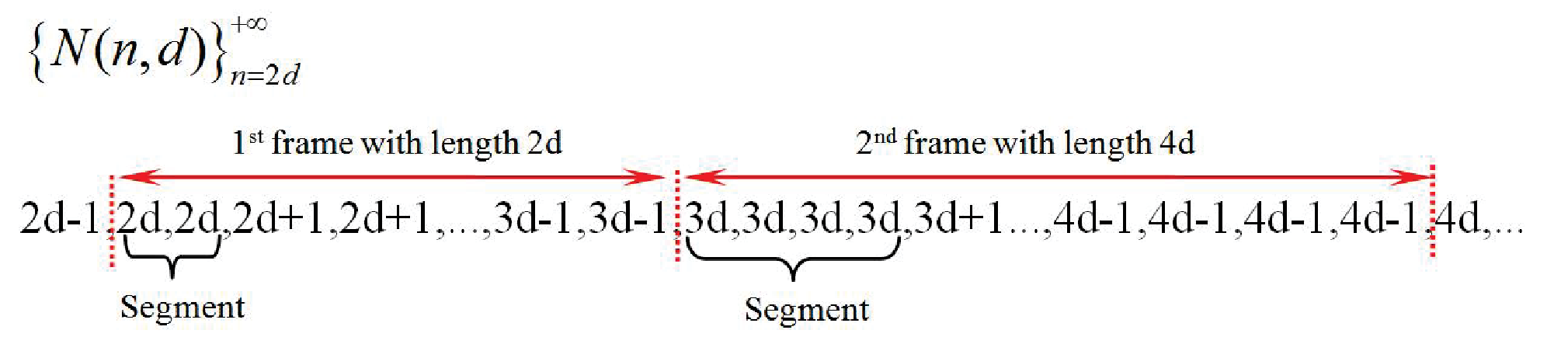}\\
\hline
\end{tabular}
\end{table}

\subsection{The Optimal Nested Test Plan}
\label{ssec:test_plan}

The theorem below characterizes the optimal nested test plan $M(n,d)$ in closed form for all $n$ and $d$.
\medskip 
\begin{theorem}\label{thm:main_pattern}
For a CGT problem $(n,d)$ with $d\le\frac{n}{2}$, we have
\begin{eqnarray}
&&\label{eqM} M(n,d)=n-2^l(d+k-1),
\end{eqnarray}
where $l$ and $k$ are the frame and segment indexes as given in \eqref{eql} and \eqref{eqk}. For $d>\frac{n}{2}$, we have
\begin{equation}
M(n,d) = M(n, n-d).
\end{equation}
\end{theorem}
\begin{table}
\centering
\caption{\label{patternM}The Frame-Segment Structure of $M(n,d)$}
\begin{tabular}{l}
\hline
\hline\\
\includegraphics[scale=0.23]{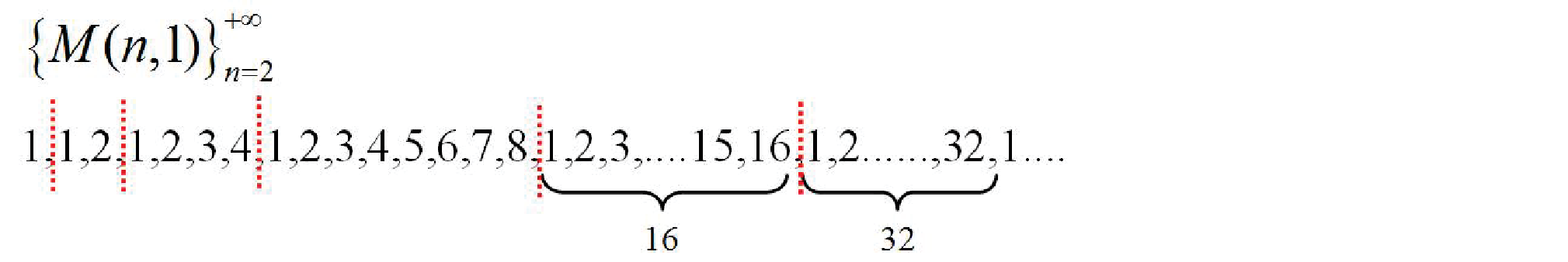}\\
\includegraphics[scale=0.23]{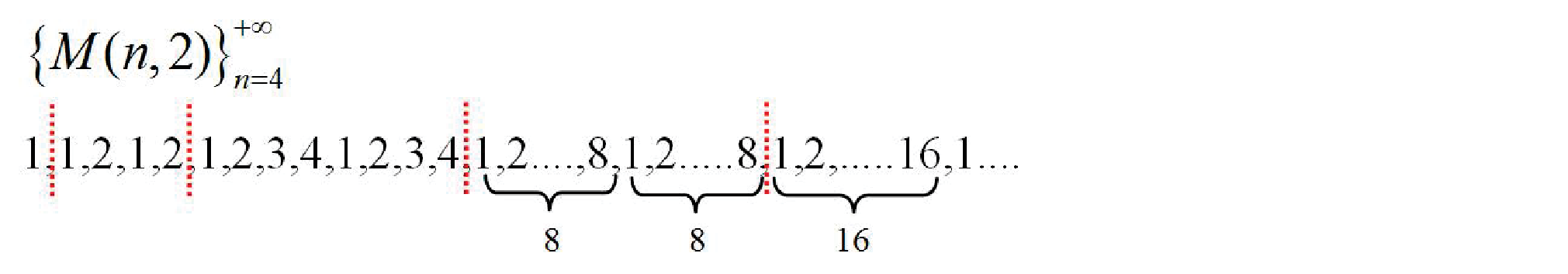}\\
$\quad\quad\vdots$\\
\includegraphics[scale=0.23]{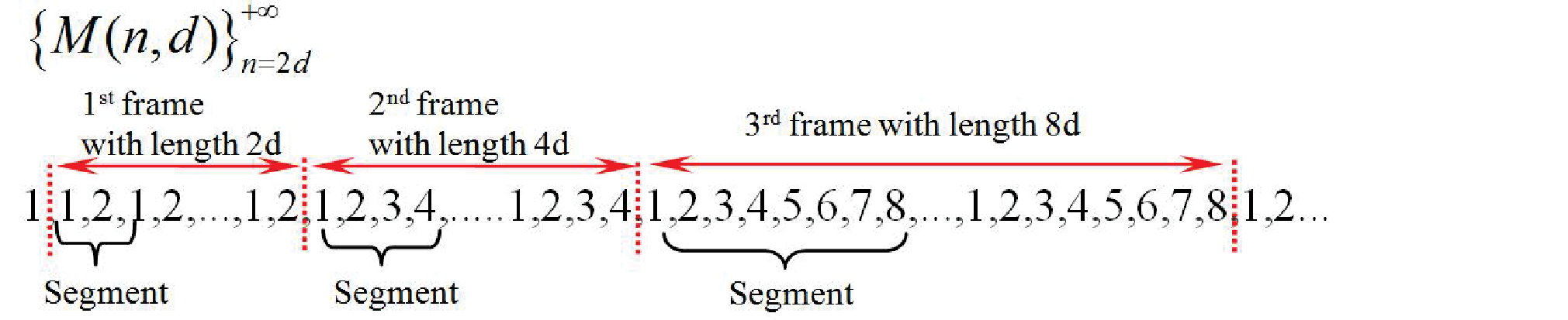}\\
\hline
\end{tabular}
\end{table}

The theorem above fully specifies the optimal nested test plan. A pseudo code implementation with a recursively called subroutine is given below.  
\begin{algorithm}
\label{algorithm}
\caption{Optimal Nested Test Plan}
\begin{algorithmic}[1]
\Require {$[n]$: a group of $n$ items;}\newline {$\text{\quad\quad}d$: number of defectives in the group.}
\Ensure $\cD$: the set of defective items.
\Procedure{Test}{$[x]$}
  \State\Return{The number of defective items in set $[x]$.}
\EndProcedure

\Procedure{Nested}{$[n],d$}
  \State{\textbf{Initialize:} $\cD = \emptyset$}
  \If{$d=0$}
    \State\Return{$\cD = \emptyset$}
  \ElsIf{$d=n$}
  	\State\Return{$\cD = [n]$}
  \Else
  	\State $[M(n,d)]$ = a subset of $[n]$ with size $M(n,d)$
  	\State $d_1$ = \Call{Test}{$[M(n,d)]$}
  	\State $\cD$ = $\cD\ \cup$ \Call{Nested}{$[M(n,d)],d_1$}
  	\State $\cD$ = $\cD\ \cup$ \Call{Nested}{$[n]\setminus [M(n,d)],d-d_1$}
  	\State\Return{$\cD$}
  \EndIf
\EndProcedure
\end{algorithmic}
\end{algorithm}

$M(n,d)$ as a sequence of $n$ for a fixed $d$ has the same frame-segment structure as $N(n,d)$. Specifically, each sequence starts at $n=2d$ with $M(2d,d)=1$. The values of $M(n,d)$ in each segment of the $l$th frame are consecutive integers from $1$ to~$2^l$. \par

We postpone the proof of Theorem~\ref{thm:main_pattern} to Section~\ref{sec:proof_thm} where we establish several key properties of $N(n,d)$ that will be used in the proof. 

\subsection{The Optimal Nested Test Plan for CGT with Unknown $d$}
\label{ssec:unknown}
We have so far focused on the standard CGT formulation which assumes a prior knowledge on the total number of defective items in the given population. For applications where this prior knowledge is unavailable, the question is how to start the first test: for any population size $n$, should the first test be carried over the entire population or a proper subset of the population with the size potentially depending on $n$? The answer is given in the following theorem.
\medskip
\begin{theorem}\label{thm:unidentified}
For a CGT problem with a population size $n$ and an unknown number of defective items, the optimal nested test plan first tests the entire population. 
\end{theorem}

\begin{IEEEproof}
Let $d$ denote the number of defective items in the population of $n$. Suppose that the first test is not carried over the entire population, but rather on a subset of $n_1$ items. Due to the nested structure, any nested test plan $\pi$ will break the problem with an unknown $d$ into a sequence of CGT problems $(n_k, d_k)$  ($k=1,2,\ldots, K$) for some  integers $K>0$, $\{n_k\}_{k=1}^K$ with $\sum_k n_k =n$, and $\{d_k\}_{k=1}^K$ with $\sum_k d_k =d$. Specifically, the test plan first tests a group of size $n_1$, and with one test revealing the number $d_1$ of defective items in this group, the test plan then resolves the CGT problem $(n_1,d_1)$. Subsequently, the test plan determines the size $n_2$ of the next group of unidentified items to test, where the choice of $n_2$ may depend on the outcomes of past tests. The procedure continues until all items are identified. We thus have
$$N_\pi(n) = K+\sum_{k=1}^K N_\pi (n_k, d_k).$$

Now consider the CGT problem $(n,d)$. A slight modification of $\pi$ that omits the group test of the last set of $n_K$ unidentified items (since the number of defective items in this last set can be deduced from past tests when $d$ is known) gives a valid nested test plan for the CGT problem $(n,d)$. 
We thus have 
$$N_\pi(n)\ge N(n,d)+1.$$
We then arrive at Theorem~\ref{thm:unidentified} by noticing that the lower bound of $N(n,d)+1$ can be achieved by first testing the entire population and that $\pi$ is an arbitrary nested test plan.
\end{IEEEproof}

With the first test revealing the total number $d$ of defective items, the problem is then reduced to a CGT of $(n,d)$.

\subsection{Order Optimality and the Approximation Ratio of the Optimal Nested Test Plan}
\label{ssec:order_opt}

The logarithmic order of $N(n,d)$ in terms of $n$ can be readily seen from the closed-form expression. Specifically, we can write $N(n,d)$ in~\eqref{eqN} as
\begin{equation}\label{two_parts}
N(n,d)={\left\lceil\log_2{\frac{n}{d}}\right\rceil\cdot d}+{\left\lceil\frac{n}{2^l}\right\rceil-d-1},
\end{equation}
where $\left\lceil\frac{n}{2^l}\right\rceil-d-1$ is bounded between $0$ and $d-1$. We compare below the order of $N(n,d)$ with that of $N^*(n,d)$, the minimum number of tests achievable among all test plans.\par

Likening the group testing problem $(n,d)$ to a source coding problem with the entropy of the source given by $\log_d\binom{n}{d}$ and each test outcome representing one letter in the corresponding codeword, we can easily obtain a lower bound of $\log_d\binom{n}{d}$ (the minimum expected codeword length) on $N^*(n,d)$. 
Thus, for all fixed $d$, the optimal nested test plan has a constant (i.e., independent of $n$) approximation ratio that is asymptotically bounded by
\begin{equation}\label{asy_ratio}
\lim_{n\rightarrow\infty} 
\frac{N(n,d)}{N^*(n,d)}\leq \log_2 d.
\end{equation}
In other words, for all fixed $d$, the optimal nested test plan is order optimal among all test plans.

Note that whether the information-theoretic lower bound of~$\log_d\binom{n}{d}$ is achievable is still an open question, since not every coding scheme can be mapped to a valid test plan. In particular, while a source code has no constraint in choosing each letter of a codeword, the sequence of test outcomes are bound by the specific configuration of the given population. For example, a test outcome cannot take a value greater than the size of the tested group, and the test outcome of a subset of a previously tested group must be consistent with the test outcome of that group. In fact, a negative answer has been established when we restrict to non-adaptive test plans. Thus, the asymptotic bound on the approximation ratio given in~\eqref{asy_ratio} may be a pessimistic one. A more detailed discussion on achievable performance and a comparison between quantitative and Boolean group testing are given in Section~\ref{sec:discussion}.

\section{Properties of $N(n,d)$ and Proof of Theorem~\ref{thm:main_pattern}}
\label{sec:proof_thm}

\subsection{Properties of $N(n,d)$}
\label{ssec:pf_property}
We first establish three properties of $N(n,d)$, which will be used in proving the closed form of $M(n,d)$ in Theorem~\ref{thm:main_pattern}. \par

\emph{Properties:}
\begin{enumerate}
\renewcommand{\labelenumi}{[P\theenumi]}
\setcounter{enumi}{0}
\item\label{p_strict} $\lbrace N(n,d)\rbrace_{d=0}^{\lfloor n/2\rfloor}$ is a strictly increasing sequence in $d$, i.e., 
$$N(n,d)>N(n,d-1),\quad \forall 1\leq d \leq \left\lfloor \frac{n}{2}\right\rfloor.$$
\item\label{p_concave} $\lbrace N(n,d)\rbrace_{d=0}^n$ is a concave sequence in $d$, i.e., for all~$1\leq d\leq n-1$, we have
$$N(n,d+1)-N(n,d)\leq N(n,d)-N(n,d-1).$$
\item\label{p_max}
For all $d\leq \left\lfloor \frac{n}{2}\right\rfloor$ and $m\leq \left\lfloor \frac{n}{2}\right\rfloor $, if 
$$N(m,0)+N(n-m,d)\geq N(m,1)+N(n-m,d-1),$$
then for all $d_1=1, 2, \ldots, \min\lbrace m,\,d\rbrace$,
\begin{equation*}
N(m,0)+N(n-m,d)\geq N(m,d_1)+N(n-m,d-d_1).
\end{equation*}
\end{enumerate}

The strict increasing property [P\ref{p_strict}] is proved via induction in $n$ and is the key property used to prove [P\ref{p_concave}]. [P\ref{p_max}] is proved based on [P\ref{p_concave}] and is the main tool for proving Theorem~\ref{thm:main_pattern}. It is used to show that, when $m=M(n,d)$, the worst case occurs at~$d_1 = 0$, i.e., the maximization over $d_1$ in~\eqref{optM} is achieved at~$d_1=0$. The proof of these three properties can be found in Appendix~\ref{app:property}.

\subsection{Proof of Theorem~\ref{thm:main_pattern}}
\label{ssec:proof_1}

We now provide a proof of Theorem~\ref{thm:main_pattern}. It suffices to consider~$d\leq\frac{n}{2}$. The proof hinges on [P\ref{p_max}] shows that when $m=M(n,d)$ the worst case occurs at $d_1 = 0$. Therefore, $N(n,d)$ equals $1 + N(n-M(n,d),d)$, which is the number of tests in the previous segment plus $1$ according to the frame-segment structure of $N(n,d)$. The detailed proof follows below.\par

We first establish the initial value $M(2d,d)=1$ of every sequence $d$. From Lemma~\ref{lemma:Nnd}, we have $N(2d,d)=2d-1$, which can be achieved by testing all but the last item one by one, i.e.,~$M(2d,d)=1$.\par

For $n>2d$, recall the frame-segment of $N(n,d)$ as illustrated in TABLE~\ref{patternM}. Consider the $x$-th ($x=1,\ldots,2^l$) element in  the $k$-th segment of the $l$-th frame, i.e.,
\begin{equation*}
n=2^{l}(d+k-1)+x.
\end{equation*}
Then \eqref{eqM} is equivalent to
\begin{equation}\label{eqM_x}
M(2^{l}(d+k-1)+x,d)=x.
\end{equation}


For notational simplicity, when the test plan selects the subset with size $m$ to test, let $\phi(m;\, n,d)$ denote the worst case number of tests for the subsequent testing under the optimal nested test plan, i.e., 
\begin{equation}\label{phi_def}
\phi(m;\, n,d) = \max_{d_1}\left\lbrace N(m,d_1) +N(n-m,d-d_1)\right\rbrace. 
\end{equation}

Recall that $M(n,d)$ is chosen as the minimum value of the group size $m$ that achieves the optimal performance $N(n,d)$. To show~\eqref{eqM_x}, it suffices to show that 
\begin{equation}
1+\phi(m;\, n,d)  
\begin{cases}
> N(n,d) & \text{when}\ m<x,\cr
= N(n,d) & \text{when}\ m=x.
\end{cases}
\end{equation}

When $m < x$, we have 
\begin{equation*}
\begin{split}
1&+\phi(m;\,2^l (d+k-1)+x, d )\\
&\overset{(a)}{\geq} 1+N(m,0)+N(2^l (d+k-1)+x-m,d)\\
&\overset{(b)}{>} N(2^l (d+k-1)+x,d),
\end{split}
\end{equation*}
where $(a)$ holds by setting $d_1=0$ in \eqref{phi_def} and $(b)$ follows from the fact that $N(2^l (d+k-1)+x-m,d) = N(2^l (d+k-1)+x,d) $ since they are in the same segment.

When $m=x$, based on Lemma~\ref{lemma:Nnd}, we have
\begin{equation*}
\begin{split}
N(x,0)&+N(2^l(d+k-1),d)-N(2^l(d+k-1),d-1)-N(x,1)\\
&=(l+1)d+k-2-(l+1)(d-1)-k+1-N(x,1)\\
&=l-N(x,1)\geq 0,
\end{split}
\end{equation*}
i.e., 
\begin{equation}\label{ineq_appd}
N(x,0)+N(2^l(d+k-1),d)\geq N(x,1)+N(2^l(d+k-1),d-1).
\end{equation}
With \eqref{ineq_appd}, based on [P\ref{p_max}],  we thus have
\begin{equation*}
\begin{split}
1&+\phi(x;\,2^l (d+k-1)+x, d)\\
&=1+\max_{d_1}\lbrace N(x,d_1)+N(2^l (d+k-1),d-d_1)\rbrace\\
&=1+N(x,0)+N(2^l (d+k-1),d)\\
&=(l+1)d+k-1\\
& = N(2^l(d+k-1)+x,d),
\end{split}
\end{equation*}
i.e., $m=x$ achieves the optimal performance~$N(2^l(d+k-1)+x,d)$. We then conclude that~$M(2^l(d+k-1)+x,d)=x$.

\section{Comparison between Quantitative and Boolean Group Testing}
\label{sec:discussion}

It is informative to summarize and compare the best known results for quantitative and Boolean CGT. In particular, it is of interest to examine the potential gain offered by quantitative test outcomes over Boolean test outcomes.

\subsection{Comparison for Cases with Known $d$}
\label{ssec:comp_known_d}
\begin{table*}
\centering
\caption{A comparative summary of boolean and quantitative group testing results.}
\label{my-label}
\begin{tabular}{|c|c|c|c|c|}
\hline
\multirow{2}{*}{} & \multicolumn{2}{c|}{Non-adaptive Test Plans} &  \multicolumn{2}{c|}{Adaptive Test Plans}\\
\cline{2-5} 
         & Lower Bound          & Upper Bound        & Lower Bound          & Upper Bound          \\ 
\hline
Boolean CGT        & $\frac{d^2\log_2 n}{24\log_2 d}$~\cite{Du_Hwang_book} & $\frac{4d^2\log^2_2 n}{\log^2_2 (d\log_2 n)}$\cite{Kautz} &$\log_2 \binom{n}{d}$ & $\log_2 \binom{n}{d} + d$~\cite{Du_Hwang_book}\\
\hline
Quantitative CGT  & $2d\log_d \left(\frac{n}{d}\right)$~\cite{D75,L75} &  $4d\log_d \left(\frac{n}{d}\right)$~\cite{GK00} (Non-constructive)  &$\log_d \binom{n}{d}$ & ${\left\lceil\log_2{\frac{n}{d}}\right\rceil\cdot d}+{\left\lceil\frac{n}{2^l}\right\rceil-d-1}$~[this work]\\ 

\hline
\end{tabular}
\label{table:bounds}
\end{table*}


We first consider the case when the total number $d$ of defective items is known. We summarize in TABLE~\ref{table:bounds} the best known lower bounds and upper bounds for Boolean CGT and quantitative CGT. \par

For Boolean CGT, when restricted to non-adaptive test plans, the tightest lower bound on the number of required tests was established in \cite{Du_Hwang_book} to be $\frac{d^2\log_2 n}{24\log_2 d}$, which is strictly greater than the information-theoretic lower bound of $\log_2 \binom{n}{d}$. In other words, the information-theoretic lower bound cannot be achieved by non-adaptive test plans. The best known non-adaptive test plan appears to be the one developed in \cite{Kautz} based on disjunct code. However, there remains a gap between the performance of this best known test plan and the tightest lower bound (see Table~III). This gap has also been studied in~\cite{aldridge2014group} from the perspective of the \emph{asymptotic rate} of the Bollean group testing algorithms. When considering adaptive test plans, the information-theoretic lower bound can be asymptotically achieved by the adaptive Generalized Binary Splitting~(GBS) algorithm developed in~\cite{Du_Hwang_book} for all fixed $d$.

For quantitative CGT, the tightest lower bound of the non-adaptive test plan on the number of required tests is $2d\log_d \left(\frac{n}{d}\right)$~\cite{L75,D75}, which is about twice the information-theoretic lower bound of $\log_d \binom{n}{d}$ for large $n$.  Grebinski and Kucherov~\cite{GK00} established the \emph{existence} of a non-adaptive test plan with a performance of $4d\log_d \left(\frac{n}{d}\right)$. However, this upper bound result is non-constructive, and no non-adaptive test plan is known to achieve this upper bound. For the adaptive test plans, the optimal nested test plan established in this work appears to be the first for the general quantitative CGT problem and achieves order optimality for all fixed $d$.

The comparison in TABLE~\ref{table:bounds} shows that results on quantitative CGT are much less complete than Boolean CGT. In particular, it remains to be an open question whether the information-theoretic lower bound can be achieved by an adaptive test plan for quantitative CGT. Consequently, whether a gain of $\log_2 d$ indicated by the information-theoretic lower bound for quantitative CGT over Booelan CGT can be realized remains elusive. Nonetheless, it can be shown that the worst-case performance of the optimal nested test plan is strictly better than the Boolean CGT lower bound $\log_2 \binom{n}{d}$. In Fig.~\ref{fig:fix_n_d_known}, we compare the average performance of the optimal nested test plan with quantitative test outcomes with that of GBS, the best known adaptive test plan for Boolean CGT. The objective is to illustrate the potential gain offered by quantitative test outcomes over Boolean test outcomes. As shown in Fig.~\ref{fig:fix_n_d_known}, for a CGT with $n=500$, the gain increases with $d$ and can be up to $25\%$.

\subsection{Comparison for Cases with Unknown $d$}
\label{ssec:comp_unknow_d}

We now consider the case with unknown $d$. For quantitative CGT, a single test of the entire population reveals $d$ and reduces the problem to a CGT $(n,d)$ with a known $d$. For Boolean CGT, however, most existing test plans rely on the knowledge of $d$ and do not easily extend to the case with unknown $d$. For example, the afore-mentioned best known non-adaptive test plan and the best adaptive test plan GBS both require the knowledge of $d$. How to estimate $d$ based on Boolean test outcomes is highly nontrivial. \par
   \begin{figure}[thpb]
      \centering
      \includegraphics[scale=0.55]{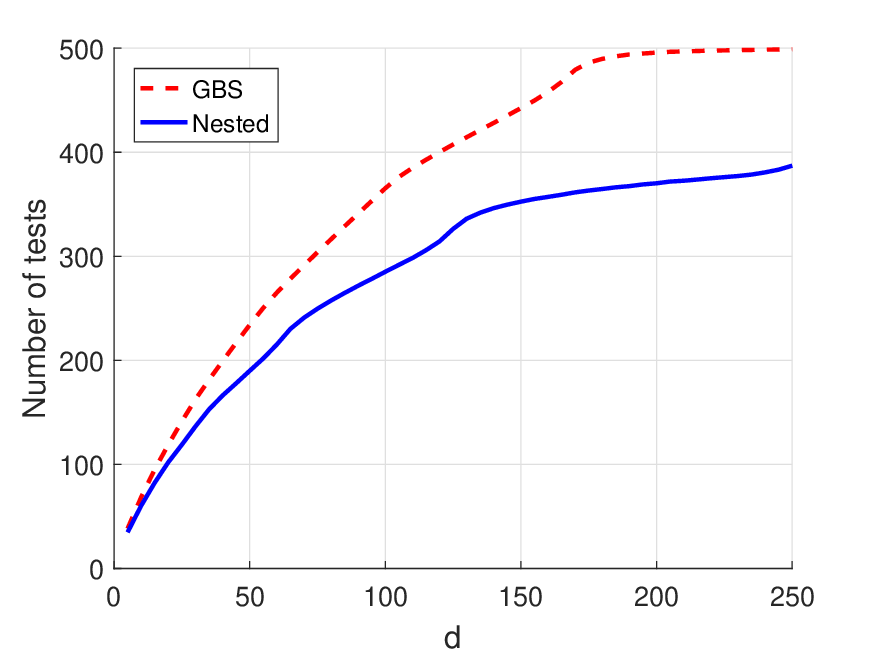}
      \caption{Comparison of the optimal nested test plan to the generalized binary splitting (GBS) test plan with known $d$ ($n=500$, $1000$ Monte Carlo runs).}
      \label{fig:fix_n_d_known}
   \end{figure}

   \begin{figure}[thpb]
      \centering
      \includegraphics[scale=0.55]{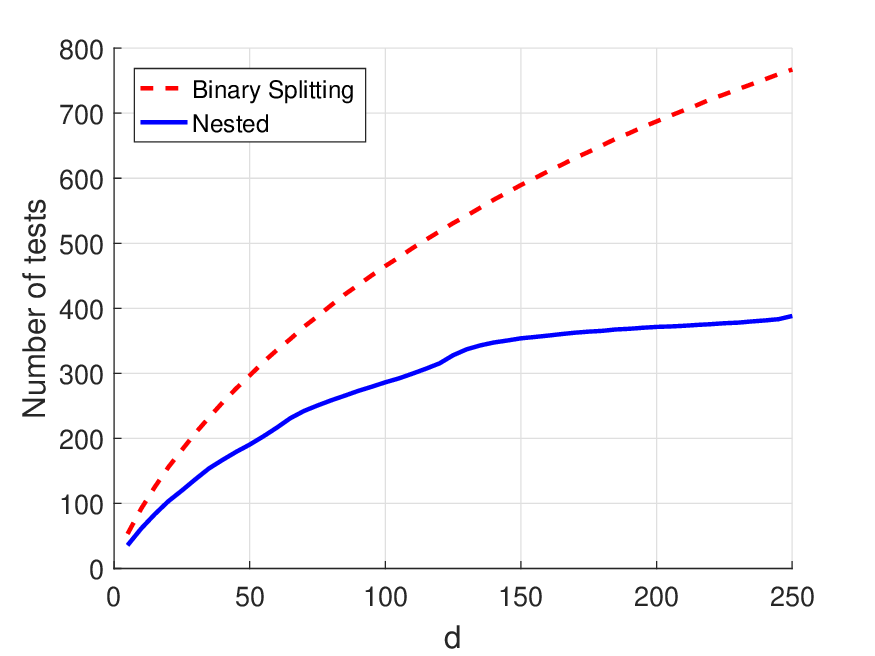}
      \caption{Comparison of the optimal nested test plan to the binary splitting test plan with unknown $d$ ($n=500$, $1000$ Monte Carlo runs).}
      \label{fig:fix_n_d_unknown}
   \end{figure}

One approach to Boolean CGT with unknown $d$ is binary splitting, which is also asymptotically optimal. In Fig.~\ref{fig:fix_n_d_unknown}, we compare the average performance of the nested test plan with quantitative test outcomes with that of a bisection search for Boolean CGT. Fig.~\ref{fig:fix_n_d_unknown} shows that when $d$ is unknown, the gain offerred by quantitative test outcomes over Boolean test outcomes increases, with up to $50\%$ gain for the same CGT problem tested in Fig.~\ref{fig:fix_n_d_known}.

\section{Application to Heavy Hitter Detection}
\label{sec:estimation}

In this section, we study the application of quantitative group testing to the heavy hitter detection problem. \par
Consider a network consisting of $n$ flows, each modeled as a random process with a certain packet arrival rate. Assume that among the $n$ flows, $n_x$ are heavy hitters with rate $\lambda_x$, and $n-n_x$ are normal flows with rate $\lambda_y$. Define 
\begin{eqnarray}
&&\label{define_rho}\rho = \frac{n_x}{n},\\
&&\label{fraction}\eta=\frac{n_x\lambda_x}{n_x\lambda_x+(n-n_x)\lambda_y}
\end{eqnarray}
as the fraction of heavy hitters in terms of the number of flows and the total traffic volume, respectively. For Internet traffic, we typically have $\rho$ around $10\%$ to $20\%$ and $\eta$ around $80\%$ to $90\%$.\par
   \begin{figure*}[thpb]
      \centering
      {\includegraphics[scale=0.55]{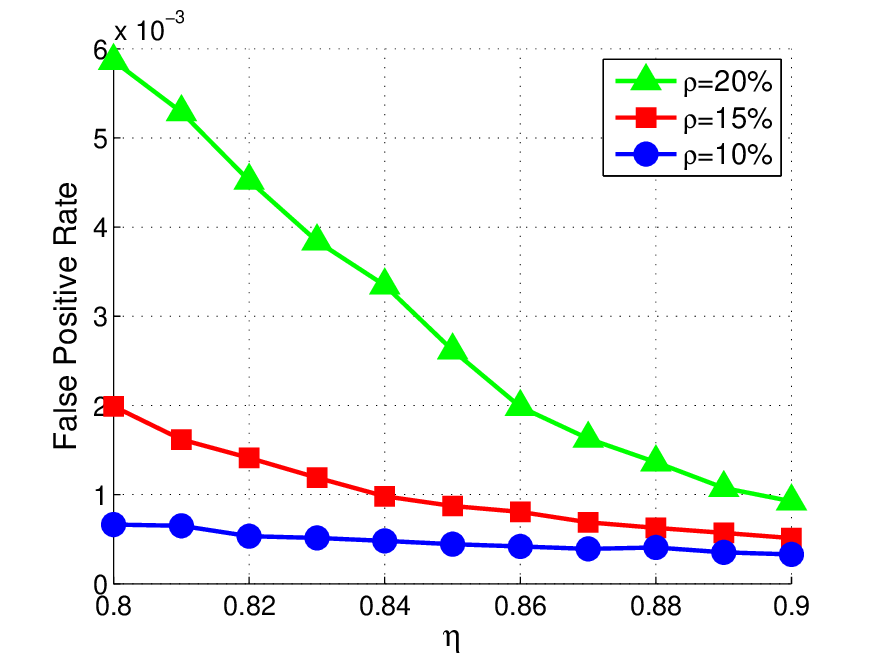}}
      {\includegraphics[scale=0.55]{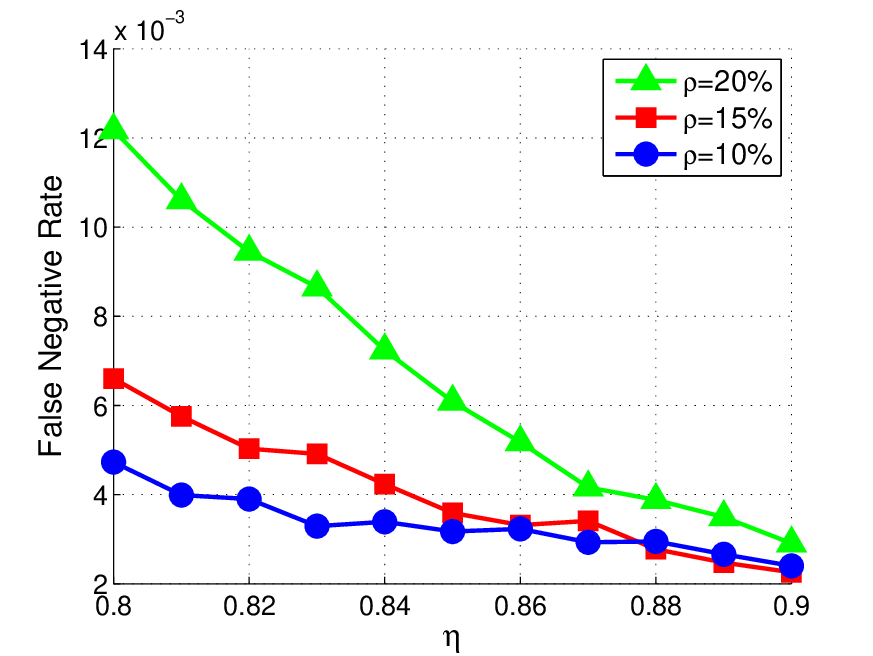}}
      \caption{Detection accuracy of the optimal nested test plan with MLE for Poisson distributed flows ($n=1000$, $T=2$, $\lambda_y=1$).}
      \label{fig:poi_detection_MLE_T2}
   \end{figure*}

   \begin{figure*}[thpb]
      \centering
      {\includegraphics[scale=0.55]{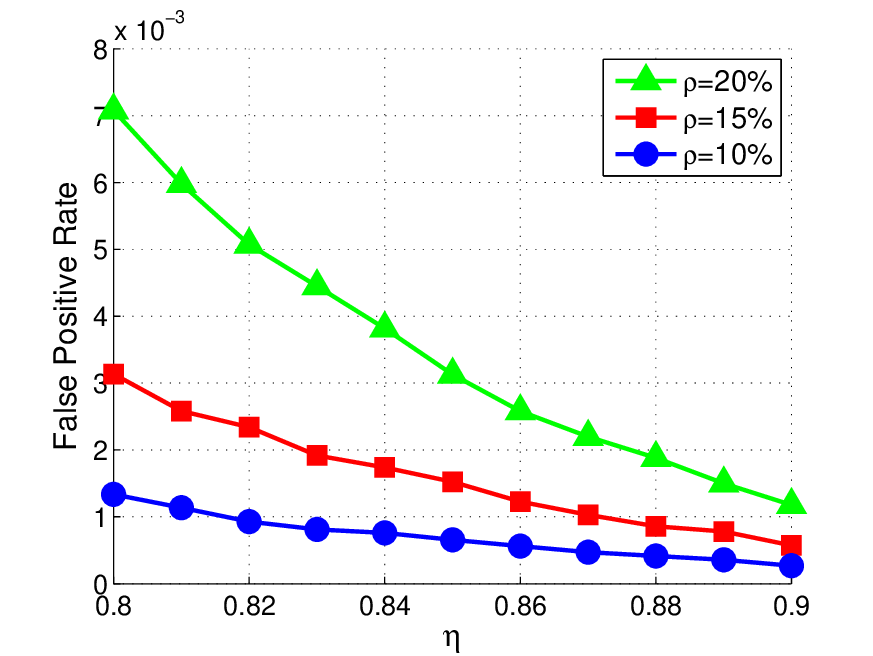}}
      {\includegraphics[scale=0.55]{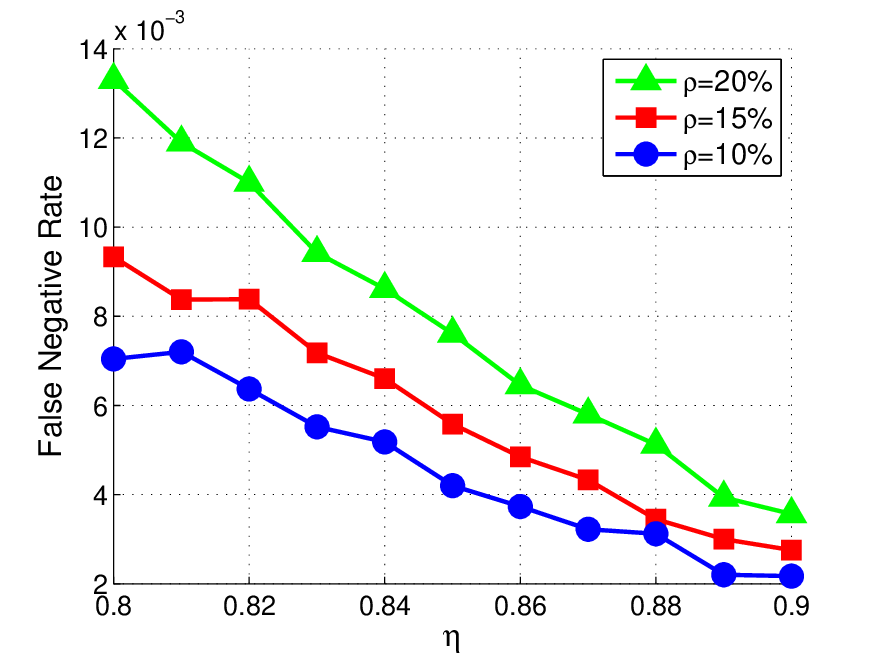}}
      \caption{Detection accuracy of the optimal nested test plan with SME for log-normal distributed flows ($n=1000$, $T=5$, $\lambda_y=1$, $\sigma_x^2=\sigma_y^2=10$).}
      \label{fig:logn_detection_RSME_T5}
   \end{figure*}
The problem is to identify the $n_x$ heavy hitters quickly and reliably. The performance metrics of interest are detection delay, detection accuracy, and counter consumption. Detection delay is defined as the average time taken to identify all heavy hitters. Detection accuracy is measured by the false positive rate $\alpha$ and false negative rate $\beta$ defined as
\begin{eqnarray}
&& \alpha = \frac{\text{Number of falsely identified heavy hitters}}{n-n_x},\\
&& \beta = \frac{\text{Number of missed heavy hitters}}{n_x}.
\end{eqnarray}
Counter consumption is given by the number of flow counters required by a heavy hitter detector. In the group testing algorithm, each test requires a counter, and the counter can be reused. Since flow counters rely on the high-speed TCAM (ternary content-addressable memory) entries which are scarce resources in routers, detectors with low counter consumption are desired.  \par

Without loss of generality, the arrival rate $\lambda_y$ of normal flows in all simulation examples is normalized to $1$. The time unit is thus determined by the expected inter-arrival time of a normal flow, which is in the millisecond scale or smaller in typical Internet traffic.

\subsection{Quantitative Group Testing for Heavy Hitter Detection}
\label{ssec:GT_HHdetection}

In the quantitative group testing formulation, it is assumed that the test result reveals the number of defective items without any error. A test plan can thus correctly identify all defective items. In the application of heavy hitter detection, the number of heavy hitters needs to be estimated from random observations of packet arrivals in an aggregated flow. The estimation errors lead to false positives and false negatives in the final detection result. We show below via simulation examples that the large gap in the arrival rates of normal flows and heavy hitters allow accurate estimation of the number of heavy hitters from random packet arrivals. Consequently, the optimal nested test plan given in Theorem~\ref{thm:main_pattern} offers attractive performance in detection accuracy. \par

In the first example, we assume that each flow is an independent Possion process. We employ the maximum likelihood estimator (MLE) in estimating the number of heavy hitters in each group test. Consider, without loss of generality, the first group test that aggregates all $n$ flows. Let $z$ denote the number of packet arrivals observed in $T$ time units in the aggregated flow. It is easy to see that the likelihood function is given by
\begin{equation*}
\begin{split}
L(n_x|z)=z\log[(n\lambda_y&+n_x(\lambda_x-\lambda_y))T]\\
&-(n\lambda_y+n_x(\lambda_x-\lambda_y))T - \log (z!).
\end{split}
\end{equation*}
The ML estimate of $n_x$ is given by
\begin{equation}\label{Real_ML}
\hat{n}_x =  \arg\max_{n_x=0,1,\ldots,n} L(n_x|z).
\end{equation}
The above integer optimization can be simplified to the following
\begin{equation}\label{Real_ML}
\hat{n}_x =  \arg\max_{n_x=i_0,i_0+1} L(n_x|z),
\end{equation}
where $i_0=\left\lfloor\frac{(z/T)-n\lambda_y}{\lambda_x-\lambda_y}\right\rfloor$. The above simplification results from the fact that $L(n_x|z)$, when viewed as a function of a real-valued argument $n_x$, is unimodal with the maximum value achieved at $\frac{(z/T)-n\lambda_y}{\lambda_x-\lambda_y}$.

From Fig.~\ref{fig:poi_detection_MLE_T2} we observe that for all typical values of $\rho$ and $\eta$, the group testing approach offers good detection reliability using only $T=2$ time units for each group test. Furthermore, the detection performance improves when $\eta$ increases and/or $\rho$ decreases, since both result in a larger gap between $\lambda_x$ and $\lambda_y$, thus better estimates of the number of heavy hitters from random packet arrivals. \par

The observation that a larger gap between the rates of heavy hitters and normal flows leads to better detection accuracy may also be deduced from the Cram\'{e}r-Rao lower bound on the mean-squared error (MSE) of estimating $n_x$. Treating $n_x$ as a real-valued argument, we obtain the lower bound as

\begin{equation}
\text{Var}(\hat{n}_x) \geq \frac{n\lambda_y +(\lambda_x-\lambda_y)n_x}{T(\lambda_x-\lambda_y)^2},
\end{equation}
showing smaller estimation error when $(\lambda_x-\lambda_y)$ increases for a fixed $\lambda_y$. Since the likelihood function is unimodal, we may expect that the MSE in estimating a real-valued proxy of $n_x$ preserves the general property of the original integer estimation problem. \par

The MLE requires the knowledge of the flow distribution and can be computationally expensive for general distributions. An alternative is a simple sample mean estimator (SME) given by
\begin{equation}\label{sample_nx}
\hat{n}_x=
\left[ \frac{z/T - n\lambda_y}{\lambda_x-\lambda_y}\right],
\end{equation}
where $[\cdot]$ denotes the operation of taking the nearest integer.\par
The detection performance of the optimal nested test plan with SME for log-normal distributed flows is shown in Fig.~\ref{fig:logn_detection_RSME_T5}. By increasing the observation time to $T=5$ for each group test, SME leads to similar detection accuracy for heavy-tailed flows. \par

   \begin{figure*}[thpb]
      \centering
      \subfigure{\includegraphics[scale=0.55]{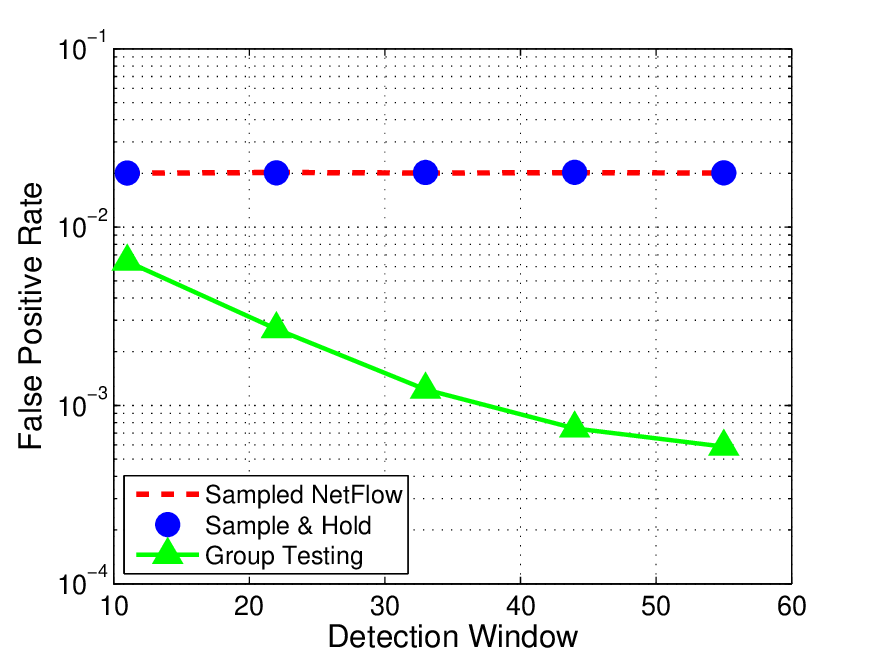}}
      \subfigure{\includegraphics[scale=0.55]{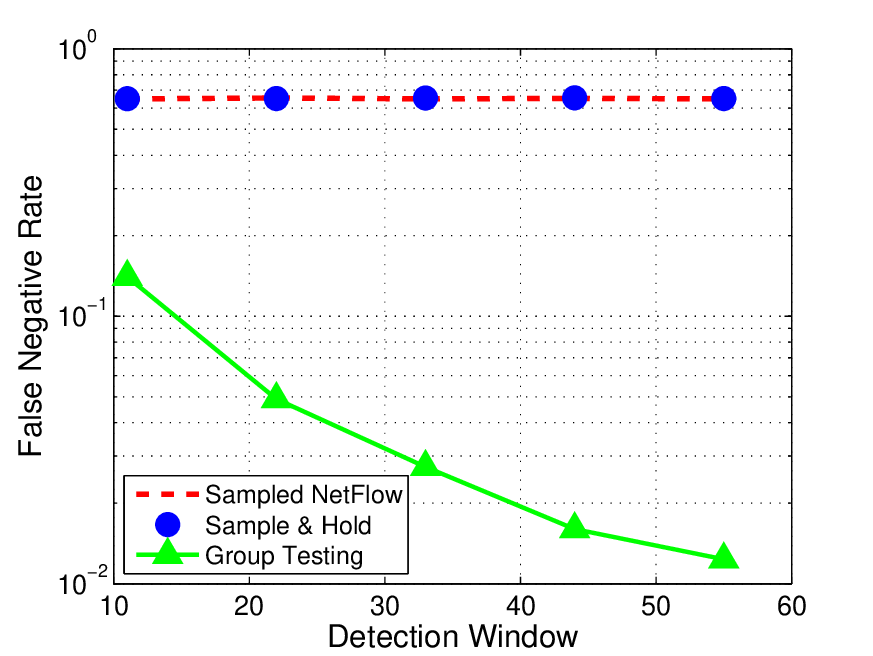}}
      \caption{Performance comparison: detection accuracy versus detection delay ($n=100$ Poisson flows, $n_x=3$, $\lambda_x=20$, $\lambda_y=1$, $c=3$).}
      \label{fig:compare_d3}
   \end{figure*}
   \begin{figure*}[thpb]
      \centering
      \subfigure{\includegraphics[scale=0.3]{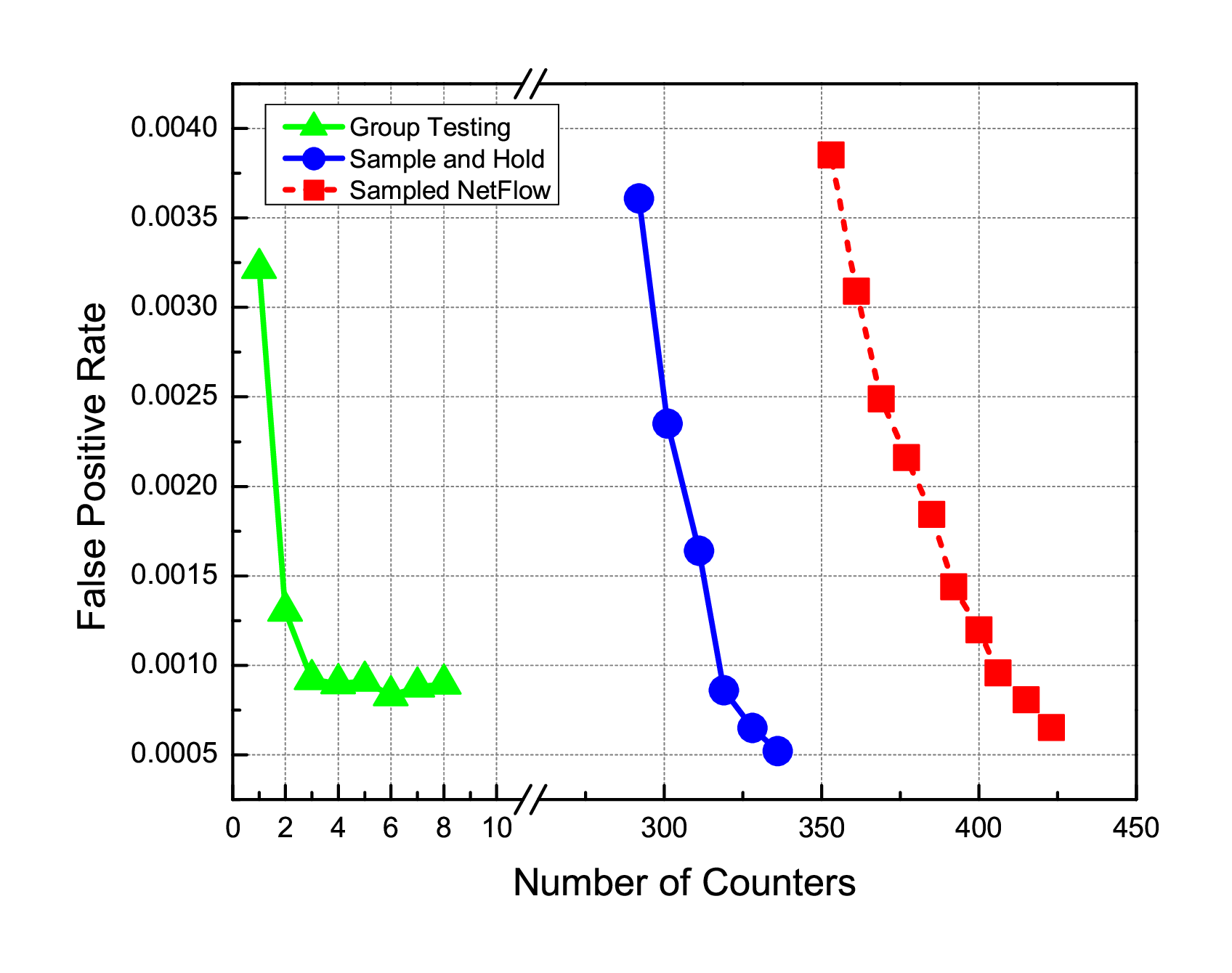}}
      \subfigure{\includegraphics[scale=0.3]{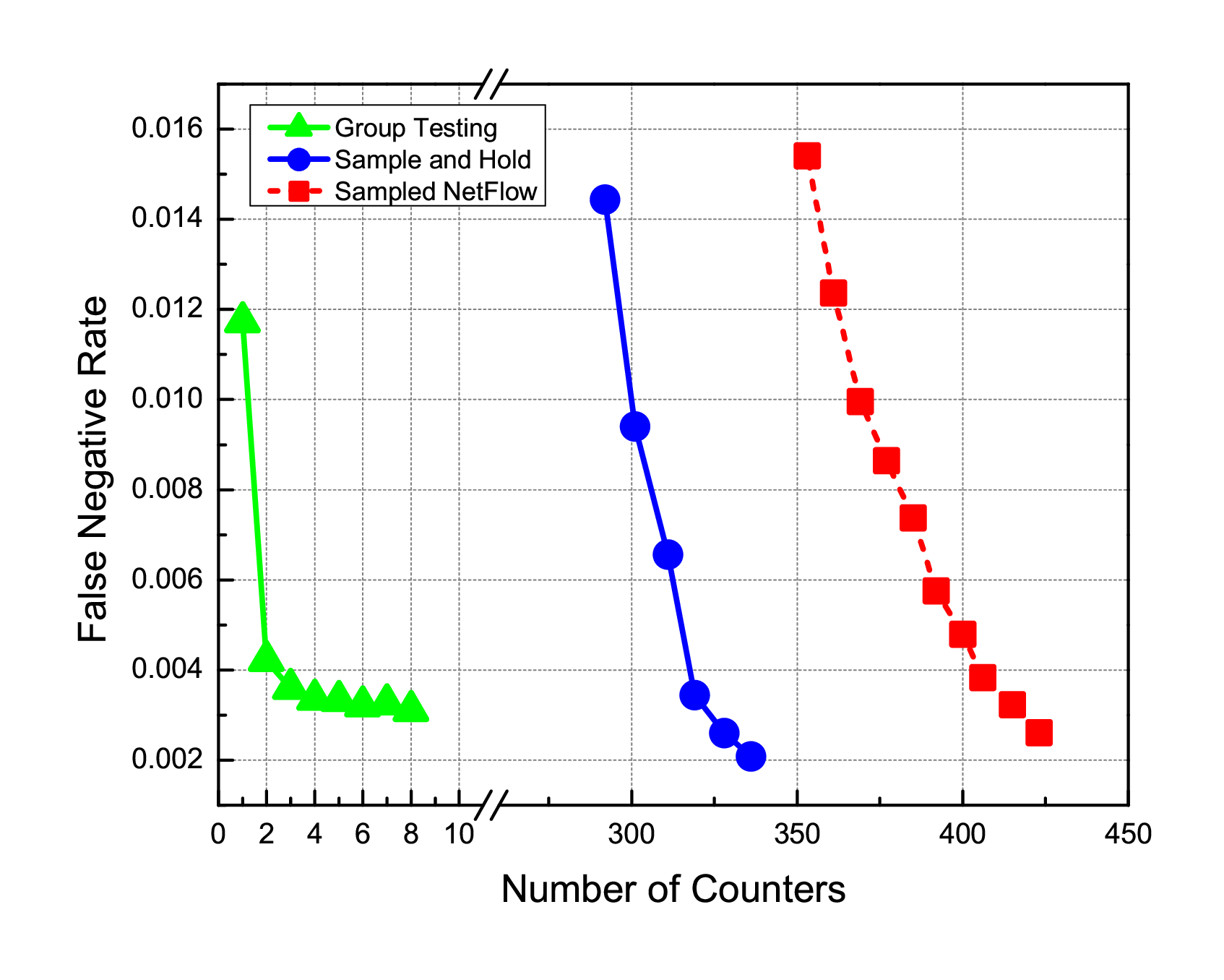}}
      \caption{Performance comparison: detection accuracy versus counter budget ($n=1000$ Poisson flows, $n_x=200$, $\lambda_x=36$, $\lambda_y=1$, $\tau=568$).}
      \label{fig:compare_d200}
   \end{figure*}

\subsection{Comparisons with Prevailing Heavy Hitter Detectors}
In this section, we compare the proposed group testing approach with two prevailing sampling-based algorithms for heavy hitter detection. The first is the Sampled NetFlow algorithm introduced and implemented by Cisco~\cite{Sanf}. Under this algorithm, one out of every $r$ packets is sampled. If the sampled packet is from a flow that has a counter established, the counter of this flow increases by one. Otherwise, a new counter is created for this flow until all available counters have been used. The sampling rate $r$ can be chosen, often heuristically, based on the router configuration. The second algorithm is the Sample and Hold scheme introduced in~\cite{Estan}. Under this algorithm, the flow ID of every packet is checked. If the packet is from a flow that has a counter established, the counter of this flow increases by one. Otherwise, with probability $p$ a new counter is created for this flow until all available counters have been used. For both algorithms, at the end of the detection window, the $n_x$ flows with the top packet counts are declared as heavy hitters, and the rest as normal flows. \par

In the first example, we compare the detection accuracy as a function of the detection window of all three algorithms under a stringent counter budget. Specifically, the total number $c$ of available counters is set to $3$. For the group testing approach, $c$ determines the maximum number of group tests that can be performed simultaneously since each group test requires counting the number of packet arrivals within an observation window of length $T$. The observation window $T$ varies from $1$ to $5$, resulting in a detection delay (i.e., detection window) of $11$ to $55$ (see the x-axis of Fig.~\ref{fig:compare_d3}). All three algorithms are implemented over the same detection window with the same realizations of the flow processes. The parameters $r$ and $p$ for the two sampling-based approaches are set to their optimal values using a brute force numerical search. From Fig.~\ref{fig:compare_d3} we observe that the group testing approach offers orders of magnitude improvement in detection accuracy under the same counter budget. Furthermore, the reliability of the group testing approach improves significantly when the detection window increases, while the reliability of the two sampling-based approaches remain roughly the same.  This is due to the fact that a longer detection window allows a longer observation window $T$ for each group test, thus smaller error in estimating the number of heavy hitters in each test. For the sampling-based approaches, however, detection accuracy is mainly limited by the counter budget. \par

In the second example, we compare the counter consumption of the three algorithms by plotting the false positive and false negative rates as functions of the number of counters as shown in Fig.~\ref{fig:compare_d200}. The detection window $\tau$ is fixed to $568$ time units for all algorithms. Again, the parameters $r$ and $p$ for the two sampling-based approaches are chosen optimally for each setting. For the group testing approach, the observation window $T$ is chosen based on the counter budget so that all tests can be finished within the detection window. More specifically, with more counters, more tests can be performed simultaneously, and each test can use more observations, resulting in better detection accuracy. In particular, with a single counter, we need to set $T=1$ in order to finish the test plan within the detection window. From Fig.~\ref{fig:compare_d200} we observe that the group testing approach reduces counter consumption from hundreds to only a handful for the same level of detection accuracy.


\section{Conclusion}
\label{sec:conclusion}
We studied the quantitative group testing problem within the combinatorial group testing framework. The optimal nested test plan was established in closed form. Its application in heavy hitter detection was studied and its performance compared with prevailing sampling-based approaches.

Besides Boolean and quantitative group testing, there are several other group testing models, including the semi-quantitative group testing(SQGT)~\cite{Emad2014semiquantitative} and the threshold group testing(TGT)~\cite{TGT}, that have been considered in the literature. Establishing the optimal nested test plan under these group testing models are interesting future directions. Since SQGT and TGT can be considered as noisy group testing with specific noise models, a recent result in~\cite{wang2017active} on a tree-structured adaptive strategy  for general noisy group testing may be applicable, even though the special noise models in SQGT and TGT may be exploited for better performance.



\appendices

\renewcommand{\thesubsection}{\arabic{subsection}}

\section{Proof of Lemma~\ref{lemma:Nnd}}
\label{app:lemma1}

Consider first the initial value of each sequence with $n=2d$. This corresponds to $l=0$, $k=d$.  
Setting $t=2$, $i=0$ in \eqref{ineq2}, we have
$$N(2d,d)\leq 2d-1.$$
Together with \eqref{ineq1}, we arrive at 
$$N(2d,d)=(l+1)d+k-1=2d-1.$$

When $n>2d$, based on the definition of $l$ and $k$, we write~$n$ as 
\begin{equation}\label{split_n}
n= (d+k-1)2^l + (n-(d+k-1)2^l).
\end{equation}
Let $x=n-(d+k-1)2^l$. It is easy to see that  $l\geq 1$, $1\leq k \leq d$, and $1\leq x \leq 2^l$. Based on~\eqref{split_n}, \eqref{eqN} is equivalent to 
\begin{equation}\label{new_Nnd}
N((d+k-1)2^l+x, d) = (l+1)d+k-1. 
\end{equation}
Setting $t=l+1$, $i=k-1$ in \eqref{ineq3}, we have
\begin{equation}\label{begin_seg}
N((d+k-1)2^l+1, d) \geq (l+1)d+k-1.
\end{equation}
Setting $t=l+1$, $i=k$ for $1\leq k\leq d-1$ and $t=l+2$, $i=0$ for $k=d$ in \eqref{ineq2}, we have,
\begin{equation}\label{end_seg}
N((d+k-1)2^l+2^l, d) \leq (l+1)d+k-1.
\end{equation}

From~\eqref{begin_seg} and \eqref{end_seg} we have, for all~$x= 1,2,\ldots,2^{t-1}$, 
\begin{equation*}
\begin{split}
(l+1)d+k-1&\leq N((d+k-1)2^l+1, d) \\
&\leq N((d+k-1)2^l+x, d) \\
&\leq N((d+k-1)2^l+2^l, d)\\
& \leq (l+1)d+k-1,
\end{split}
\end{equation*}
which leads to \eqref{new_Nnd}. Here we have used the monotonicity property of $N(n,d)$, i.e., $N(n,d)\leq N(n+1,d), \forall n\geq 2d$.

\section{Proof of Properties of $N(n,d)$}
\label{app:property}

In this appendix, we provide the proof of the three properties [P\ref{p_strict}]-[P\ref{p_max}] introduced in Section~\ref{sec:proof_thm}.

\subsection{Proof of [P\ref{p_strict}]}
\label{app:pf_strict}

The proof is based on induction in $n$ using the recursive formulas in \eqref{opt} and \eqref{optM}. 

Let~$d_1^{*}(m;\, n,d)$ denote the maximizer that achieves $\phi(m;\, n,d)$ as defined in \eqref{phi_def}, i.e.,
\begin{equation}\label{phi_opt}
\phi(m;\, n,d)=N(m,d_1^{*}(m;\, n,d))+N(n-m,d-d_1^{*}(m;\, n,d)).
\end{equation}
Note that since $d$ and $m$ are restricted to no greater than $\frac{n}{2}$, we have $0\leq d_1^{*}(m;\, n_0,d)\leq \min\{m,d\}$.

The initial condition of the induction is easy to check: $N(2,1)=1>N(2,0)=0$. Now assume that there exists an $n_0>2$ such that for every $n<n_0$, $\lbrace N(n,d)\rbrace_{d=0}^{\lfloor n/2\rfloor}$ is a strictly increasing sequence in $d$. Based on this induction assumption, we prove next that $\lbrace N(n_0,d)\rbrace_{d=0}^{\lfloor n_0/2\rfloor}$ is strictly increasing in~$d$.\par
It is straightforward that $N(n_0,0)<N(n_0,1)$. 
When $d>2$, we prove the statement by considering separately the cases when $n_0$ is odd and when $n_0$ is even.\\ 
\emph{Case 1:} $n_0$ is odd.\par

The basic idea of the proof is to show that for all~$m = 1,\ldots,\left\lfloor \frac{n}{2}\right \rfloor$, 
\begin{equation}\label{phi_ineq}
\phi(m;\, n_0,d-1)< \phi(m;\, n_0,d).
\end{equation}
Then from \eqref{opt}, we arrive at [P\ref{p_strict}].\par
Next, we show \eqref{phi_ineq} by considering the following two cases in terms of the value of $d_1^*(m;\, n_0,d-1)$:
\begin{align}
\label{cond1}&0\leq d_1^*(m;\, n_0,d-1)< \min\left\lbrace  \left\lfloor \frac{m}{2} \right\rfloor , d \right\rbrace,\\
\label{cond2}&d-1-\left\lfloor \frac{n_0-m}{2} \right\rfloor < d_1^*(m;\, n_0,d-1) \leq \min\{m,\,d-1\}. 
\end{align}
It is easy to see that \eqref{cond1} and \eqref{cond2} cover all possible values of $d_1^*(m;\, n_0,d-1)$ since the upper limit in \eqref{cond1} is greater than the lower limit in \eqref{cond2} given that $m\leq \left\lfloor \frac{n_0}{2} \right\rfloor$ and $d\leq \left\lfloor \frac{n_0}{2} \right\rfloor$.\par
When \eqref{cond1} is true,  we have
\begin{equation*}\label{ieq1}
\begin{split}
\phi(&m; \, n_0,d)\\
& 
\begin{split}
\overset{(a)}{\geq} N(m,d_1^{*}(m;\, &n_0,d-1)+1)\\
&+N(n_0-m,d-d_1^{*}(m;n_0,d-1)-1)
\end{split}\\
&
\begin{split}
\overset{(b)}{>}N(m,d_1^{*}(m;\, &n_0,d-1))\\
&+N(n_0-m,d-d_1^{*}(m;n_0,d-1)-1)
\end{split}\\
&\overset{(c)}{=}\phi(m; \, n_0,d-1),
\end{split}
\end{equation*}
where $(a)$ holds since $d_1^{*}(m;\, n_0,d-1)+1$ is in the range $\{0, \ldots, \min\{m,d\}\}$ of the maximizer for $\phi(m;\, n_0,d)$; $(b)$ follows from the induction hypothesis and the fact that $d_1^{*}(m;\, n_0,d-1) < \left\lfloor \frac{m}{2}\right\rfloor$ given in \eqref{cond1}, and $(c)$ follows from \eqref{phi_opt}. We thus arrive at~\eqref{phi_ineq}. \par

When \eqref{cond2} is true, 
by noticing that $d_1^{*}(m;\, n_0,d-1)$ is within the range $\{0, \ldots, \min\{m,d\}\}$ of the maximizer for $\phi(m; \, n_0,d)$, we have
\begin{equation*}\label{ieq2}
\begin{split}
\phi(&m; \, n_0,d)\\
& 
\begin{split}
{\geq} N(m,d_1^{*}(m;\, &n_0,d-1))\\
&+N(n_0-m,d-d_1^{*}(m;n_0,d-1))
\end{split}\\
&
\begin{split}
{>}N(m,d_1^{*}(m;\, &n_0,d-1))\\
&+N(n_0-m,d-d_1^{*}(m;n_0,d-1)-1)
\end{split}\\
&{=}\phi(m; \, n_0,d-1).
\end{split}
\end{equation*}
This concludes the proof for \emph{Case 1}. 

\emph{Case 2:} $n_0$ is even. \par

For $d<\frac{n_0}{2}$, the proof follows the same line of argument as in \emph{Case 1}. Now consider $d=\frac{n_0}{2}$. We need to prove $N(n_0,\frac{n_0}{2}-1)<N(n_0,\frac{n_0}{2})$. Base on Lemma~\ref{lemma:Nnd}, we have $N(n_0,\frac{n_0}{2})=n_0-1$. Then it is equivalent to prove~$N(n_0,\frac{n_0}{2}-1)< n_0-1$. \par 
When $m$ is even, $d_1^*(m;\, n_0,\frac{n_0}{2}-1)$ is covered by \eqref{cond1} and \eqref{cond2}. The same line of arguments as in \emph{Case 1} leads to
\begin{equation}
\phi(m;\, n_0,\frac{n_0}{2}-1)< \phi(m;\, n_0,\frac{n_0}{2}).
\end{equation}
Based on the unimodal property of $\lbrace N(n,d)\rbrace_{d=0}^n$ given in [P3], we further have
\begin{equation}
d_1^{*}(m;\, n_0,\frac{n_0}{2}) = \frac{m}{2}, 
\end{equation}
i.e., 
\begin{align}
\phi(m;\, n_0,\frac{n_0}{2})&= N(m, \frac{m}{2} )+N(n_0-m, \frac{n_0-m}{2})\\
\label{eq_L1}&=n_0-2,
\end{align}
where \eqref{eq_L1} is based on Lemma~\ref{lemma:Nnd}.
Therefore, we have, for all even $m$,
\begin{equation}\label{even_m}
\phi(m;\, n_0,\frac{n_0}{2}-1)<  n_0-2. 
\end{equation}

When $m$ is odd, based on [P3] and Lemma~\ref{lemma:Nnd}, we have
\begin{equation}\label{odd_m}
\begin{split}
\phi(m;\, n_0, \frac{n_0}{2} -1) &= N(m, \frac{m-1}{2} )+N(n_0-m, \frac{n_0-m-1}{2})\\
&=n_0-2.
\end{split}
\end{equation}
With \eqref{even_m} and \eqref{odd_m}, we have
\begin{equation*}
N(n_0,\frac{n_0}{2}-1)=1+\min_{m}\phi(m;\, n_0,\frac{n_0}{2}-1) < n_0-1,
\end{equation*}
i.e.,
\begin{equation}
N(n_0,\frac{n_0}{2}-1) < N(n_0,\frac{n_0}{2}).
\end{equation}

\subsection{Proof of [P\ref{p_concave}]}\label{app:pf_concave}
We first establish the following lemma. 
\begin{lemma}\label{lemma:concave}
Let $f(x)$ be a real-valued function defined on a finite set of consecutive integers, i.e., $x\in\{u,u+1,\ldots, v\}$ for some $u$ and $v$. Suppose that $f(x)$ is monotonically increasing and concave.  For every positive integer $s$, let $\lbrace c_k\rbrace_{k=0}^s$ be an arbitrary increasing and concave sequence. Define, for $x=u, u+1, \ldots, v+s$,
\begin{equation*}
F(x):= \max \lbrace f(x)+c_0, f(x-1)+c_1, \ldots, f(x-\tau)+c_\tau\rbrace, 
\end{equation*}
where $\tau=\min\{x-u, s\}$. 
Then $F(x)$ is increasing and concave.
\end{lemma}

\begin{figure}[thpb]
      \centering
      \includegraphics[scale=0.55]{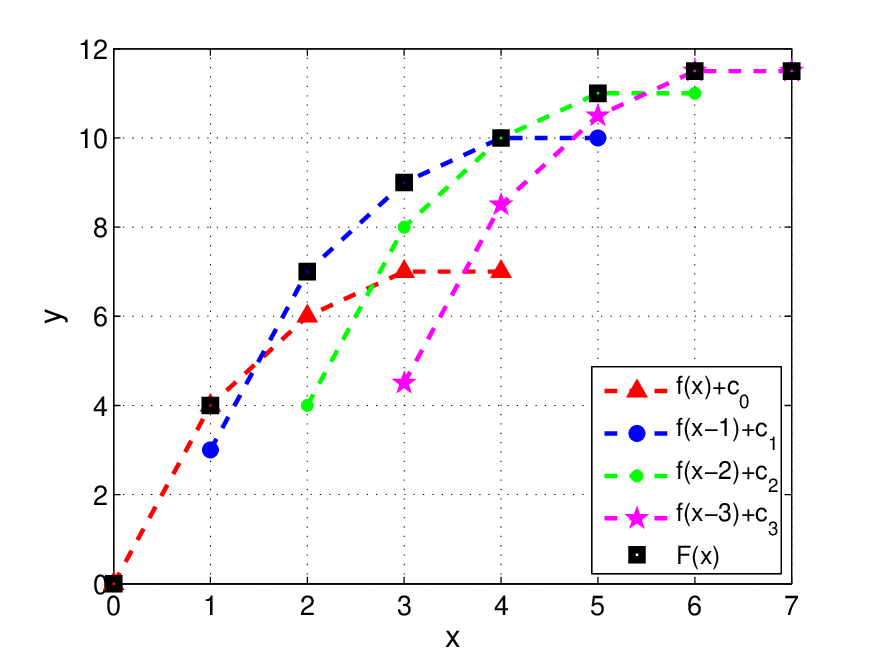}
      \caption{Illustration of Lemma~\ref{lemma:concave} with $s=3$.}
      \label{fig:concave4}
   \end{figure}

This lemma is rather intuitive given that $F(x)$ is the maximum of shifted versions of $f(x)$ which is increasing and concave. An numerical example with $s=3$ is given in Fig.~\ref{fig:concave4}.\par
Next we provide a detailed proof of this lemma. 
Our objective is to show
\begin{equation}\label{obj}
0\leq F(x+1)-F(x) \leq F(x) - F(x-1).
\end{equation}
Based on the definition of $F(x)$, we have
\begin{equation*}
\begin{split}
F(x-1) &= \max_{k=0,1,\ldots, \min\{x-u-1, s\}}\left\lbrace f(x-k-1)+c_k \right\rbrace,\\
F(x) &= \max_{k=0,1,\ldots, \min\{x-u, s\}}\left\lbrace f(x-k)+c_k\right\rbrace,\\
F(x+1) &= \max_{k=0,1,\ldots, \min\{x-u+1, s\}}\left\lbrace f(x-k+1)+c_k \right\rbrace.
\end{split}
\end{equation*}
Define $k^*$ as  
$$k^* = \arg\max_{k=0,1,\ldots,\min\{x-u, s\}}\left\lbrace f(x-k)+c_k\right\rbrace,$$
i.e.,
$$F(x) = f(x-k^*) + c_{k^*}.$$
Based on this definition, we have
\begin{equation}\label{assump1}
f(x-k^*)-f(x-(k^*+1)) \geq c_{k^*+1}-c_{k^*},
\end{equation}
\begin{equation}\label{assump2}
f(x-(k^*-1))-f(x-k^*) \leq c_{k^*}-c_{k^*-1}.
\end{equation}
Based on~\eqref{assump1} and~\eqref{assump2}, due to the monotonicity and the concavity of $f(x)$ and $\lbrace c_k\rbrace_{k=1}^{n}$, we can easily prove
\begin{equation}\label{conv_prop_geq}
\begin{split}
f(x-(k^*+1))- &f(x-(k^*+2))\\ 
&\geq f(x-k^*)-f(x-(k^*+1)) \\
&\geq c_{k^*+1}-c_{k^*} \\
&\geq c_{k^*+2}-c_{k^*+1}
\end{split}
\end{equation}
\begin{equation}\label{conv_prop_leq}
\begin{split}
f(x-(k^*-2))-&f(x-(k^*-1)) \\
&\leq f(x-(k^*-1))-f(x-k^*) \\
&\leq c_{k^*}-c_{k^*-1} \\
&\leq c_{k^*-1}-c_{k^*-2}
\end{split}
\end{equation}
From \eqref{conv_prop_geq} and \eqref{conv_prop_leq}, respectively, we have
\begin{equation}\label{next_geq}
f(x-(k^*+1))+c_{k^*+1} \geq f(x-(k^*+2))+c_{k^*+2},
\end{equation}
\begin{equation}\label{before_leq}
f(x-(k^*-2))+c_{k^*-2} \leq f(x-(k^*-1))+c_{k^*-1}.
\end{equation}

\emph{Remark}: Based on the same arguments introduced above, it is not difficult to prove that $f(x-k)+c_k$ is monotonically increasing in $k$ when $k\leq k^*$ and monotonically decreasing in $k$ when $k\geq k^*$. \par

Next, based on the definition of $k^*$ and the above remark, we will prove that
\begin{equation}\label{fx-1}
F(x-1) = \max\left\lbrace f(x-k^*) +c_{k^*-1}, f(x-k^*-1) +c_{k^*}\right\rbrace,
\end{equation}
\begin{equation}\label{fx+1}
F(x+1) = \max\left\lbrace f(x-k^*+1) +c_{k^*}, f(x-k^*) +c_{k^*+1}\right\rbrace.
\end{equation}
Notice that, here we assume $f(x-k-1)$ and $c_{k+1}$ have definitions on $k=k^*$. If either of them does not exist, without loss of generality, we can simply discard the corresponding item in the maximum equation in~\eqref{fx-1} and~\eqref{fx+1}.\par
Based on~\eqref{conv_prop_geq}, we have 
\begin{equation}
f(x-(k^*+1))-f(x-(k^*+2))  \geq c_{k^*+1}-c_{k^*},
\end{equation}
i.e.,
\begin{equation}\label{left_x-1}
f(x-(k^*+1))-c_{k^*} \geq f(x-(k^*+2)) + c_{k^*+1}.
\end{equation}

Based on~\eqref{conv_prop_leq}, we have 
\begin{equation}
f(x-(k^*-1))-f(x-k^*) \leq c_{k^*-1}-c_{k^*-2},
\end{equation}
i.e.,
\begin{equation}\label{right_x-1}
f(x-(k^*-1))+ c_{k^*-2}\leq f(x-k^*)+c_{k^*-1}.
\end{equation}

Due to~\eqref{left_x-1},~\eqref{right_x-1} and the previous remark, we can shown that $f(x-k-1)+c_k$ is monotonically increasing with $k$ when $k\leq k^*-1$ and monotonically decreasing with $k$ when $k\geq k^*$. \eqref{fx-1} is thus proved. 

Similarly, based on \eqref{conv_prop_geq} and \eqref{conv_prop_leq}, we can show that
\begin{equation}
\begin{split}
&f(x-k^*) + c_{k^*+1} \geq f(x-(k^*+1)) + c_{k^*+2},\\
&f(x-(k^*-2)) +c_{k^*-1} \leq f(x-(k^*-1)) +c_{k^*},
\end{split}
\end{equation}
which shows that $f(x-k+1)+c_k$ is monotonically increasing with $k$ when $k\leq k^*$ and monotonically decreasing with $k$ when $k\geq k^*+1$ based on the above remark. \eqref{fx+1} is thus proved. \par

In conclusion, now we have 
\begin{equation*}
\begin{split}
F(x-1) &= \max\left\lbrace f(x-k^*) +c_{k^*-1}, f(x-k^*-1) +c_{k^*}\right\rbrace,\\
F(x) &= f(x-k^*) +c_{k^*},\\
F(x+1) &= \max\left\lbrace f(x-k^*+1) +c_{k^*}, f(x-k^*) +c_{k^*+1}\right\rbrace.
\end{split}
\end{equation*}

If $F(x-1)=f(x-k^*) +c_{k^*-1}$ and $F(x+1)= f(x-k^*) +c_{k^*+1}$, based on the monotonicity and the concavity of $c_k$, the objective equation~\eqref{obj} is true.\par
If $F(x-1)=f(x-k^*-1) +c_{k^*}$ and $F(x+1)= f(x-k^*+1) +c_{k^*}$, based on the monotonicity and the concavity of $f(x)$, the objective equation~\eqref{obj} is true. \par
If $F(x-1)=f(x-k^*) +c_{k^*-1}$ and $F(x+1)= f(x-k^*+1) +c_{k^*}$, we have 
\begin{equation}
\begin{split}
F(x) - F(x-1) &= c_{k^*}-c_{k^*-1} \geq 0,\\
F(x+1) - F(x) &= f(x-k^*+1) - f(x-k^*) \geq 0.
\end{split}
\end{equation}
Then, based on~\eqref{assump2}, the objective equation~\eqref{obj} is proved. \par
If $F(x-1)=f(x-k^*-1) +c_{k^*}$ and $F(x+1)= f(x-k^*) +c_{k^*+1}$, we have 
\begin{equation}
\begin{split}
F(x) - F(x-1) &= f(x-k^*) - f(x-k^*-1) \geq 0,\\
F(x+1) - F(x) &= c_{k^*+1}-c_{k^*} \geq 0.
\end{split}
\end{equation}
Then, based on~\eqref{assump1}, we arrive at~\eqref{obj}, which completes the proof of Lemma~\ref{lemma:concave}. 

\medskip

We now prove [P\ref{p_concave}] based on Lemma~\ref{lemma:concave}.  Based on the symmetry property of $N(n,d)$, it is sufficient to consider $d=0,1,\ldots, \left\lfloor \frac{n}{2} \right\rfloor$. The proof is based on induction in $n$ using the recursive formulas in (3,\,4). The initial condition of the induction is easy to check: $N(2,d)$ is a concave function of $d$ for $0\leq d\leq 1$. Now assume that there exists an $n_0>2$ such that for every $n<n_0$, $\lbrace N(n,d)\rbrace_{d=0}^{\lfloor n/2 \rfloor}$ is a concave sequence in $d$. \par
Based on this induction assumption, we prove next that $\lbrace N(n_0,d)\rbrace_{d=0}^{\lfloor n_0/2 \rfloor}$ is a concave sequence in $d$.\par

In the following proof, for given $n$ and $m$, $\phi(m;\, n,d)$ in \eqref{phi_def} is viewed as a function of $d$. The maximizer of \eqref{phi_def} defined as
\begin{equation*}\label{d1_d}
d_1^*(m;\, n, d) :=\arg \max_{d_1}\lbrace N(m,d_1)+N(n-m,d-d_1)\rbrace
\end{equation*}
is also viewed as a function of $d$. \par

We show next that for given $n_0$ and $m$, $\phi(m;\, n_0, d)$ is concave in $d$. Since $m\leq \left\lfloor \frac{n_0}{2}\right\rfloor$, $d\leq \left\lfloor \frac{n_0}{2}\right\rfloor$, based on the symmetric property $N(n,d) = N(n,n-d)$ and the increasing property [P\ref{p_strict}], $d_1^*(m;\, n_0,d)$ must satisfy
\begin{equation}
d- \left\lceil\frac{n_0-m}{2}\right\rceil\leq d_1^*(m;\, n_0,d) \leq \left\lfloor \frac{m}{2}\right\rfloor.
\end{equation}
Also since $0 \leq d_1^*(m;\, n_0,d) \leq d$, 
we can tighten the range of $d_1^*(m;\, n_0,d)$ to
$$\underline{d_1} \leq d_1^*(m;\, n_0,d) \leq  \overline{d_1},$$
where $\underline{d_1} = \max\left\lbrace 0, d-\left\lceil\frac{n_0-m}{2}\right\rceil\right\rbrace$, $\overline{d_1}=\min\left\lbrace d,  \left\lfloor \frac{m}{2}\right\rfloor \right\rbrace$.

Thus $\phi(m;\, n_0,d)$ can be written as
\begin{equation}\label{phi_d2}
\begin{split}
\phi(m;\, n_0,d) = \max_{d_{1}=\underline{d_1},\ldots \overline{d_1}}\lbrace N(m,d_1)+N(n_0-m,d-d_1)\rbrace.
\end{split}
\end{equation}
Note that  
$N(n_0-m,d-\underline{d_1})$ is an increasing (based on [P\ref{p_strict}]) and concave (based on the induction hypothesis) function of $d$. For the same reasons, $\lbrace  N(m,{d_1})\rbrace_{d_1=\underline{d_1}}^{\overline{d_1}}$ is an increasing and concave sequence in $d_1$. Then Lemma~\ref{lemma:concave} immediately shows that $\phi(m;\, n_0,d)$ is increasing and concave in $d$, i.e.,
$$2\phi(m;\, n_0, d)\geq \phi(m;\, n_0, d-1)+\phi(m;\, n_0, d+1).$$
We thus have
\begin{equation}
\begin{split}
\min_{m}&\{2\phi(m;\, n_0, d)\}\\
&\geq \min_{m}\{\phi(m;\, n_0, d-1)+\phi(m;\, n_0, d+1)\}\\
&\geq \min_{m}\{\phi(m;\, n_0, d-1)\}+\min_{m}\{\phi(m;\, n_0, d+1)\}.
\end{split}
\end{equation}
Adding $2$ to both sides of the inequality, we complete the induction and arrive at [P\ref{p_concave}].

\subsection{Proof of [P\ref{p_max}]}\label{app:pf_max}
[P\ref{p_max}] can be easily deduced from [P\ref{p_concave}] as follows. \par

The condition in [P\ref{p_max}] is equivalent to
$$N(m,1)-N(m,0) \leq N(n-m,d)-N(n-m,d-1).$$
Applying the concavity property in [P\ref{p_concave}] to both sides of this equality leads to
$$N(m,2)-N(m,1) \le N(n-m, d-1) - N(n-m, d-2),$$
which is equivalent to the statement in [P\ref{p_max}] for~$d_1=2$. Following the same line of argument, we arrive at [P\ref{p_max}] for~$d_1=1, 2, \ldots, \min\{m,d\}$.

\bibliographystyle{ieeetr}
\bibliography{WangZhaoChuah_TSP}

\begin{thebibliography}{10}

\bibitem{Dorfman}
R.~Dorfman, ``The detection of defective members of large populations,'' {\em
  The Annals of Mathematical Statistics}, vol.~14, no.~4, pp.~436--440, 1943.

\bibitem{sobel1959}
M.~Sobel and P.~A. Groll, ``Group testing to eliminate efficiently all
  defectives in a binomial sample,'' {\em Bell System Technical Journal},
  vol.~38, no.~5, pp.~1179--1252, 1959.

\bibitem{Li}
C.~H. Li, ``A sequential method for screening experimental variables,'' {\em
  Journal of the American Statistical Association}, vol.~57, no.~298,
  pp.~455--477, 1962.

\bibitem{Katona}
G.~O. Katona, ``Combinatorial search problems,'' {\em A survey of combinatorial
  theory}, pp.~285--308, 1973.

\bibitem{Du_Hwang_book}
D.~Du and F.~Hwang, {\em Combinatorial group testing and its applications}.
\newblock World Scientific, 2nd~ed., 2000.

\bibitem{Ngo}
H.~Q. Ngo and D.-Z. Du, ``A survey on combinatorial group testing algorithms
  with applications to {DNA} library screening,'' {\em Discrete mathematical
  problems with medical applications}, vol.~55, pp.~171--182, 2000.

\bibitem{Du_Hwang_book_nonadp}
D.~Du and F.~Hwang, {\em Pooling Design and Nonadaptive Group Testing:
  Important Tools for DNA Sequencing}.
\newblock World Scientific, 2006.

\bibitem{JKWolf}
J.~Wolf, ``Born again group testing: Multiaccess communications,'' {\em IEEE
  Transactions on Information Theory}, vol.~31, pp.~185--191, Mar 1985.

\bibitem{wolf1985principles}
J.~K. Wolf, ``Principles of group testing and an application to the design and
  analysis of multi-access protocols,'' in {\em The Impact of Processing
  Techniques on Communications}, pp.~237--257, Springer, 1985.

\bibitem{Berger}
T.~Berger, N.~Mehravari, D.~Towsley, and J.~Wolf, ``Random multiple-access
  communication and group testing,'' {\em IEEE Transactions on Communications},
  vol.~32, pp.~769--779, Jul 1984.

\bibitem{Murthy}
A.~Sharma and C.~Murthy, ``Group testing based spectrum hole search for
  cognitive radios,'' {\em IEEE Transactions on Vehicular Technology}, vol.~PP,
  no.~99, pp.~1--1, 2014.

\bibitem{Cheraghchi_comp}
M.~Cheraghchi, A.~Hormati, A.~Karbasi, and M.~Vetterli, ``Group testing with
  probabilistic tests: Theory, design and application,'' {\em IEEE Transactions
  on Information Theory}, vol.~57, pp.~7057--7067, Oct 2011.

\bibitem{Cheraghchi_graph}
M.~Cheraghchi, A.~Karbasi, S.~Mohajer, and V.~Saligrama, ``Graph-constrained
  group testing,'' {\em IEEE Transactions on Information Theory}, vol.~58,
  pp.~248--262, Jan 2012.

\bibitem{Thai}
M.~T. Thai, Y.~Xuan, I.~Shin, and T.~Znati, ``On detection of malicious users
  using group testing techniques,'' in {\em The 28th International Conference
  on Distributed Computing Systems}, pp.~206--213, 2008.

\bibitem{Khattab}
S.~Khattab, S.~Gobriel, R.~Melhem, and D.~Mosse, ``Live baiting for
  service-level {D}o{S} attackers,'' in {\em The 27th IEEE Conference on
  Computer Communications}, April 2008.

\bibitem{Balding}
D.~Balding, W.~Bruno, D.~Torney, and E.~Knill, ``A comparative survey of
  non-adaptive pooling designs,'' in {\em Genetic mapping and DNA sequencing},
  pp.~133--154, Springer, 1996.

\bibitem{Problem1399}
H.~S. Shapiro, ``Problem {E} 1399,'' {\em Amer. Math. Monthly}, vol.~67,
  no.~82, pp.~697--697, 1960.

\bibitem{F60}
N.~Fine, ``Solution of problem {E} 1399,'' {\em American Mathematical Monthly},
  vol.~67, no.~7, pp.~697--698, 1960.

\bibitem{L75}
B.~Lindstr{\"o}m {\em et~al.}, ``Determining subsets by unramified
  experiments,'' 1975.

\bibitem{ER63}
P.~Erd{\"o}s and A.~R{\'e}nyi, ``{On two problems of information theory},''
  1963.

\bibitem{D75}
A.~Djackov, ``On a search model of false coins,'' in {\em Topics in Information
  Theory (Colloquia Mathematica Societatis Janos Bolyai 16, Keszthely,
  Hungary). Budapest, Hungary: Hungarian Acad. Sci}, p.~163170, 1975.

\bibitem{CK08}
S.-S. Choi and J.~H. Kim, ``Optimal query complexity bounds for finding
  graphs,'' in {\em Proceedings of the 40th annual ACM symposium on Theory of
  computing}, pp.~749--758, ACM, 2008.

\bibitem{A86}
M.~Aigner, ``Search problems on graphs,'' {\em Discrete Applied Mathematics},
  vol.~14, no.~3, pp.~215--230, 1986.

\bibitem{Hao90}
F.~H. Hao, ``The optimal procedures for quantitative group testing,'' {\em
  Discrete Applied Mathematics}, vol.~26, no.~1, pp.~79--86, 1990.

\bibitem{GMSV92}
L.~Gargano, V.~Montouri, G.~Setaro, and U.~Vaccaro, ``An improved algorithm for
  quantitative group testing,'' {\em Discrete applied mathematics}, vol.~36,
  no.~3, pp.~299--306, 1992.

\bibitem{AS85}
M.~Aigner and M.~Schughart, ``Determining defectives in a linear order,'' {\em
  Journal of Statistical Planning and Inference}, vol.~12, pp.~359--368, 1985.

\bibitem{B09}
N.~H. Bshouty, ``Optimal algorithms for the coin weighing problem with a spring
  scale.,'' in {\em COLT}, 2009.

\bibitem{CW79}
S.-C. Chang and E.~Weldon, ``Coding for {T}-user multiple-access channels,''
  {\em IEEE Transactions on Information Theory}, vol.~25, no.~6, pp.~684--691,
  1979.

\bibitem{BM10}
N.~H. Bshouty and H.~Mazzawi, ``Toward a deterministic polynomial time
  algorithm with optimal additive query complexity,'' in {\em Mathematical
  Foundations of Computer Science}, pp.~221--232, Springer, 2010.

\bibitem{GK97}
V.~Grebinski and G.~Kucherov, ``Optimal reconstruction of graphs under the
  additive model,'' in {\em Algorithms{-}ESA'97}, pp.~246--258, Springer, 1997.

\bibitem{Han&Frazier&Jedynak:14}
W.~Han, P.~I. Frazier, and B.~M. Jedynak, ``Probabilistic group testing under
  sum observations: A parallelizable 2-approximation for entropy loss,'' {\em
  arXiv preprint arXiv:1407.4446}, 2015.

\bibitem{Thompson_wide}
K.~Thompson, G.~Miller, and R.~Wilder, ``Wide-area internet traffic patterns
  and characteristics,'' {\em IEEE Network}, vol.~11, pp.~10--23, Nov 1997.

\bibitem{Fang}
W.~Fang and L.~Peterson, ``Inter-{AS} traffic patterns and their
  implications,'' in {\em Global Telecommunications Conference}, vol.~3,
  pp.~1859--1868, 1999.

\bibitem{Yu2013SDN}
M.~Yu, L.~Jose, and R.~Miao, ``Software defined traffic measurement with
  {O}pen{S}ketch.,'' in {\em NSDI}, vol.~13, pp.~29--42, 2013.

\bibitem{Kautz1964nonrandom}
W.~Kautz and R.~Singleton, ``Nonrandom binary superimposed codes,'' {\em IEEE
  Transactions on Information Theory}, vol.~10, no.~4, pp.~363--377, 1964.

\bibitem{Berger2002asymptotic}
T.~Berger and V.~I. Levenshtein, ``Asymptotic efficiency of two-stage
  disjunctive testing,'' {\em IEEE Transactions on Information Theory},
  vol.~48, no.~7, pp.~1741--1749, 2002.

\bibitem{Gilbert2008group}
A.~C. Gilbert, M.~A. Iwen, and M.~J. Strauss, ``Group testing and sparse signal
  recovery,'' in {\em 42nd Asilomar Conference on Signals, Systems and
  Computers}, pp.~1059--1063, 2008.

\bibitem{Emad2014semiquantitative}
A.~Emad and O.~Milenkovic, ``Semiquantitative group testing,'' {\em IEEE
  Transactions on Information Theory}, vol.~60, no.~8, pp.~4614--4636, 2014.

\bibitem{Erdos1985families}
P.~Erd{\"o}s, P.~Frankl, and Z.~F{\"u}redi, ``Families of finite sets in which
  no set is covered by the union of others,'' {\em Israel Journal of
  Mathematics}, vol.~51, no.~1, pp.~79--89, 1985.

\bibitem{Xu2007_efficient}
W.~Xu and B.~Hassibi, ``Efficient compressive sensing with deterministic
  guarantees using expander graphs,'' in {\em IEEE Information Theory
  Workshop}, pp.~414--419, 2007.

\bibitem{Xu2007_further}
W.~Xu and B.~Hassibi, ``Further results on performance analysis for compressive
  sensing using expander graphs,'' in {\em 41st Asilomar Conference on Signals,
  Systems and Computers}, pp.~621--625, 2007.

\bibitem{Indyk2008_near_L1}
P.~Indyk and M.~Ruzic, ``Near-optimal sparse recovery in the l1 norm,'' in {\em
  49th Annual IEEE Symposium on Foundations of Computer Science}, pp.~199--207,
  2008.

\bibitem{Jafarpour2009_efficient}
S.~Jafarpour, W.~Xu, B.~Hassibi, and R.~Calderbank, ``Efficient and robust
  compressed sensing using optimized expander graphs,'' {\em IEEE Transactions
  on Information Theory}, vol.~55, no.~9, pp.~4299--4308, 2009.

\bibitem{Berinde2008_combining}
R.~Berinde, A.~C. Gilbert, P.~Indyk, H.~Karloff, and M.~J. Strauss, ``Combining
  geometry and combinatorics: A unified approach to sparse signal recovery,''
  in {\em 46th Annual Allerton Conference on Communication, Control, and
  Computing}, pp.~798--805, 2008.

\bibitem{cheraghchi2010derandomization}
M.~Cheraghchi, ``Derandomization and group testing,'' in {\em 48th Annual
  Allerton Conference on Communication, Control, and Computing}, pp.~991--997,
  2010.

\bibitem{Malloy2014_near_opt_adp}
M.~L. Malloy and R.~D. Nowak, ``Near-optimal adaptive compressed sensing,''
  {\em IEEE Transactions on Information Theory}, vol.~60, no.~7,
  pp.~4001--4012, 2014.

\bibitem{Iwen2012adaptive}
M.~A. Iwen and A.~H. Tewfik, ``Adaptive strategies for target detection and
  localization in noisy environments,'' {\em IEEE Transactions on Signal
  Processing}, vol.~60, no.~5, pp.~2344--2353, 2012.

\bibitem{Haupt2012sequentially}
J.~Haupt, R.~Baraniuk, R.~Castro, and R.~Nowak, ``Sequentially designed
  compressed sensing,'' in {\em IEEE Statistical Signal Processing Workshop
  (SSP)}, pp.~401--404, 2012.

\bibitem{Ji2008bayesian}
S.~Ji, Y.~Xue, and L.~Carin, ``Bayesian compressive sensing,'' {\em IEEE
  Transactions on Signal Processing}, vol.~56, no.~6, pp.~2346--2356, 2008.

\bibitem{Kautz}
W.~Kautz and R.~Singleton, ``Nonrandom binary superimposed codes,'' {\em IEEE
  Transactions on Information Theory}, vol.~10, no.~4, pp.~363--377, 1964.

\bibitem{GK00}
V.~Grebinski and G.~Kucherov, ``Optimal reconstruction of graphs under the
  additive model,'' {\em Algorithmica}, vol.~28, no.~1, pp.~104--124, 2000.

\bibitem{aldridge2014group}
M.~Aldridge, L.~Baldassini, and O.~Johnson, ``Group testing algorithms: Bounds
  and simulations,'' {\em IEEE Transactions on Information Theory}, vol.~60,
  no.~6, pp.~3671--3687, 2014.

\bibitem{Sanf}
Cisco, ``Sampled {N}et{F}low.''
  \url{http://www.cisco.com/c/en/us/td/docs/ios/12_0s/feature/guide/12s_sanf.html}.

\bibitem{Estan}
C.~Estan and G.~Varghese, {\em New directions in traffic measurement and
  accounting}, vol.~32.
\newblock ACM, 2002.

\bibitem{TGT}
P.~Damaschke, ``Threshold group testing,'' in {\em General theory of
  information transfer and combinatorics}, pp.~707--718, Springer, 2006.

\bibitem{wang2017active}
C.~Wang, K.~Cohen, and Q.~Zhao, ``Active hypothesis testing on a tree: Anomaly
  detection under hierarchical observations,'' in {\em IEEE International
  Symposium onInformation Theory (ISIT)}, pp.~993--997, 2017.

\end{thebibliography}

\end{document}